\documentclass[submission, Phys]{SciPost}

\newcommand{\eq}[1]{\begin{equation}#1\end{equation}}
\newcommand{\dd}{\mathrm{d}}
\newcommand{\ee}{\mathrm{e}}

\newcommand{\Ai}{\mathrm{Ai}}
\newcommand{\Tr}{\mathrm{Tr \,}}

\newcommand{\identity}{\mathbb{1}}
\newcommand{\trb}{\mathrm{Tr}_B}

\newcommand{\mnt}{\mathcal{M}_n(t)}
\newcommand{\cmnt}{\mathcal{C}_{m,n}(t)}
\newcommand{\rhoua}{\rho^{\Uparrow}_A}
\newcommand{\rhoda}{\rho^{\Downarrow}_A}
\newcommand{\mupr}{\ket{\Uparrow}}
\newcommand{\mupl}{\bra{\Uparrow}}
\newcommand{\mdor}{\ket{\Downarrow}}
\newcommand{\mdol}{\bra{\Downarrow}}

\newcommand{\nsr}[1]{\ket{#1}_{\mathrm{NS}}}
\newcommand{\nsl}[1]{{\vphantom{\ket{#1}}}_\mathrm{NS}{\bra{#1}}}
\newcommand{\rr}[1]{\ket{#1}_{\mathrm{R}}}
\newcommand{\rl}[1]{{\vphantom{\ket{#1}}}_\mathrm{R}{\bra{#1}}}

\usepackage{amsmath}
\usepackage{amsfonts}
\usepackage{braket}
\usepackage{bbold}
\usepackage{graphics}
\usepackage{graphicx}
\usepackage{leftidx}
\usepackage{psfrag}
\usepackage{color}
\usepackage{colordvi}
\usepackage{bbm}

\begin{document}

\begin{center}{\Large \textbf{
Front dynamics in the XY chain after local excitations
}}\end{center}

\begin{center}
Viktor Eisler\textsuperscript{},
Florian Maislinger\textsuperscript{}
\end{center}

\begin{center}
Institut f\"ur Theoretische Physik, Technische Universit\"at Graz,
Petersgasse 16, A-8010 Graz, Austria
\\
\end{center}

\begin{center}
\today
\end{center}



\section*{Abstract}
{\bf
We study the time evolution of magnetization and entanglement for initial states
with local excitations, created upon the ferromagnetic ground state of the XY chain.
For excitations corresponding to a single or two well separated domain walls, the
magnetization profile has a simple hydrodynamic limit, which has a standard
interpretation in terms of quasiparticles. In contrast, for a spin-flip we obtain an
interference term, which has to do with the nonlocality of the excitation in the
fermionic basis. Surprisingly, for the single domain wall the hydrodynamic limit
of the entropy and magnetization profiles are found to be directly related.
Furthermore, the entropy profile is additive for the double domain wall, whereas
in case of the spin-flip excitation one has a nontrivial behaviour.
}

\vspace{10pt}
\noindent\rule{\textwidth}{1pt}
\tableofcontents\thispagestyle{fancy}
\noindent\rule{\textwidth}{1pt}
\vspace{10pt}

\section{Introduction}

The nonequilibrium dynamics of integrable quantum many-body systems
has been the focus of intensive research \cite{CEM16}.
The interest in these peculiar models, characterized by the existence of a
large set of conservation laws, comes from two main perspectives.
On one hand, they show relaxation towards generalized stationary ensembles
that are not described by conventional statistical mechanics \cite{VR16}.
On the other hand, owing to the presence of stable quasiparticle excitations,
integrable models have anomalous transport properties \cite{VM16}.
A recent milestone in understanding the transport driven by an initial
inhomogeneity has been the formulation of generalized
hydrodynamics (GHD) \cite{BCDNF16,CADY16}, which gives accurate
predictions for the profiles of conserved densities in an appropriate
spacetime scaling limit.

The simplest paradigm of an inhomogeneous initial state is a domain wall,
separating domains of spins with different magnetizations.
Letting the system evolve, the domain wall starts to melt, giving rise to
an expanding front region characterized by a nonzero spin current.
The resulting magnetization profiles were studied in various integrable spin
models such as the XX chain \cite{ARRS99,AKR08,VSDH15},
the transverse Ising (TI) \cite{Karevski02,PK05,Kormos17}, the XY \cite{Lancaster16}
as well as the XXZ chains \cite{GKSS05,JZ11,SM13,BCDNF16,CDLV18,GE19}.
Rather generically one finds ballistic transport, with the exception of
the isotropic Heisenberg chain where a diffusive behaviour is observed instead
\cite{LZP17a,LZP17b,Stephan17,MMK17,MPP19,GMI19}.
The common feature in all of the examples above is that the domain wall
is oriented along the $z$-axis, and thus the magnetization is a local operator
in the fermionic representation of the corresponding spin chain.
In particular, for models with fermion-number conservation, the transverse
magnetization itself corresponds to a locally conserved density, which
makes the problem directly amenable to GHD techniques.

Recently, however, domain walls created upon the symmetry-broken
ferromagnetic ground states of TI or XY chains have been considered
\cite{ZGEN15,EME16,EM18}. The ordering in these chains occurs in
the longitudinal component of the magnetization, which is a highly nonlocal
string operator in the fermionic picture, being nontrivially related to the
local conserved densities. Hence, even though one has a free-fermion
model at hand, it is a priori unclear whether a hydrodynamic
description still holds for this observable. Nevertheless, in \cite{EME16,EM18} it has
been shown that, for domain walls excited by a single local fermion operator,
the longitudinal magnetization profile has the usual hydrodynamic scaling limit
one would naively expect. Namely, the profile is determined by noninteracting
quasiparticles carrying the fraction of a spin-flip and traveling at the
corresponding group velocity.

In the present work we extend these studies to excitations that
can be written as the product of two local fermion operators.
In the spin language they describe a double domain wall, and
if the distance between them is sufficiently large, we find that the
magnetization profile factorizes in the hydrodynamic scaling limit.
In other words, the quasiparticle excitations created
at the two domain walls are completely independent.
In contrast, the situation becomes nontrivial if the fermionic
excitations act on neighbouring sites, even though the product
of two adjacent domain walls is just a spin-flip and thus perfectly
local in the spin-representation. Indeed, it turns out that this
composite fermionic excitation leads to interference effects between
the quasiparticle modes, encoded in the form factors of the spin operator.
This interference term yields a significant contribution to the hydrodynamic
profile, which can be found analytically via stationary phase analysis.

We also study in detail the correlation functions and the entanglement entropy
for the single domain wall excitation. Interestingly, both of them can be
directly related to the magnetization. For the correlations we derive
a relation which holds also for finite times if the separation of the spins
is much larger than the correlation length. On the other hand, for the entropy
we propose an ansatz that is motivated by recent results for single-mode
quasiparticle excitations in a free massive quantum field theory (QFT)
\cite{CADFDS18a,CADFDS18b}. Our ansatz works perfectly
in the hydrodynamic regime, thereby creating an exact relation between
the magnetization and entanglement profiles. Furthermore, we observe
that the entropy becomes additive for the double domain wall excitation,
whereas for the spin-flip one has again a nontrivial behaviour due to the
above mentioned interference terms.

The paper is structured as follows. We start by introducing the model
in Sec. \ref{sec:model}. The magnetization dynamics is studied in
Sec. \ref{sec:mag} for three different local excitations as well as for a local
quench. The correlation functions are investigated in Sec. \ref{sec:corr},
followed by the study of the entropy profiles in Sec. \ref{sec:ent}.
We discuss our findings in Sec. \ref{sec:disc}, and the technical details
of the calculations are reported in three Appendices.

\section{Model\label{sec:model}}

We consider an XY spin chain of length $N$ described by the Hamiltonian
\eq{
H=- \sum_{n=1}^{N-1} \left( \frac{1+\gamma}{4}\sigma_n^x 
\sigma_{n+1}^x+\frac{1-\gamma}{4}\sigma_n^y \sigma_{n+1}^y\right)
-\frac{h}{2} \sum_{n=1}^{N} \sigma_n^z \, ,
\label{hxy}}
where $\sigma_n^\alpha$ are Pauli matrices located at site $n$,
$h$ and $\gamma$ denote the transverse magnetic field and the
XY anisotropy, respectively. We restrict ourselves to the parameter
regime $0<h<1$ and $0<\gamma \le 1$ where the chain is in a
gapped ferromagnetic phase, with $\gamma=1$ corresponding to
the TI chain.

The Hamiltonian \eqref{hxy} is diagonalized through a standard procedure \cite{FF},
by first introducing Majorana fermions via a Jordan-Wigner transformation
\eq{
a_{2j-1} = \prod_{k=1}^{j-1} \sigma^z_k \, \sigma^x_{j} \, , \qquad
a_{2j} = \prod_{k=1}^{j-1} \sigma^z_k \, \sigma^y_{j} \, ,
\label{maj}}
satisfying anticommutation relations $\left\{a_k,a_l\right\}=2\delta_{k,l}$.
While \eqref{hxy} describes an open chain which is most suitable for our
numerical calculations, the analytical treatment of the problem requires to
consider either periodic ($s=+$) or antiperiodic ($s=-$) boundary conditions,
$\sigma^x_{N+1}=s \sigma^x_1$ and $\sigma^y_{N+1}=s \sigma^y_1$.
Due to the global spin-flip symmetry of the model, the corresponding
Hamiltonians can then be split into two parts
\eq{
H_s = \frac{1-s\mathcal{P}}{2}H_{\mathrm{R}} +
\frac{1+s\mathcal{P}}{2}H_{\mathrm{NS}} \, , \qquad
\mathcal{P} = \prod_{n=1}^{N} \sigma^z_n \, .
}
In terms of the Majorana fermions, the corresponding symmetry sectors
are described by the Hamiltonians
\eq{
H_{\mathrm{R/NS}}=\frac{i}{2}\sum_{j=1}^{N} \left(
\frac{1+\gamma}{2} a_{2j}a_{2j+1}
-\frac{1-\gamma}{2} a_{2j-1}a_{2j+2}
+h a_{2j-1}a_{2j} \right),\qquad
\label{hrns}}
which differ in the boundary conditions $a_{2N+1}=\pm a_1$ and $a_{2N+2}=\pm a_2$
being periodic for the Ramond (R) and antiperiodic for the Neveu-Schwarz (NS) sectors.

In order to diagonalize \eqref{hrns}, one performs a Fourier transformation followed
by a Bogoliubov rotation
\eq{
\begin{split}
a_{2j-1}=\frac{1}{\sqrt{N}} \sum_{q \in \mathrm{R/NS}}
\ee^{-iqj} \ee^{i(\theta_q+q)/2} (b_q^\dag + b_{-q}), \\
a_{2j}=\frac{-i}{\sqrt{N}} \sum_{q \in \mathrm{R/NS}}
\ee^{-iqj} \ee^{-i(\theta_q+q)/2} (b_q^\dag - b_{-q}),
\end{split}
\label{ab}}
where the Bogoliubov angle and the dispersion are given by
\eq{
\ee^{i(\theta_q+q)} = \frac{\cos q -h + i \gamma \sin q}{\epsilon_q}, \qquad
\epsilon_q = \sqrt{(\cos q-h)^2+\gamma^2 \sin^2 q} \, .
\label{epsq}}
Note that the above definition ensures that the function $\theta_q$ is
continuous within the Brillouin zone $q \in \left[-\pi, \pi\right]$.
To satisfy the proper boundary conditions, the allowed values of the momenta
are $q_k=\frac{2\pi}{N}k$ for R and $q_k=\frac{2\pi}{N}(k+1/2)$ for NS, respectively,
with $k=-N/2,\dots,N/2-1$ and $N$ even. The diagonalized Hamiltonian and its $K$-particle
eigenstates are then given by
\eq{
H_{\mathrm{R/NS}} = \sum_{q \in \mathrm{R/NS}}
{\epsilon_q} b^\dag_q b_q + \mathrm{const}, \qquad
|q_1,q_2,\dots,q_K\rangle_{\mathrm{R/NS}} = 
\prod_{i=1}^{K} b^\dag_{q_i} |0\rangle_{\mathrm{R/NS}} \, .
\label{hdiag}}
It should be stressed that the eigenstates with $K$ even belong to the
spin-periodic Hamiltonian $H_+$, whereas the eigenstates of the
spin-antiperiodic $H_-$ have odd $K$.

In the thermodynamic limit $N\to \infty$, the periodic chain $H_+$ has a doubly
degenerate ground state with ferromagnetic ordering along the $x$-axis, denoted by
$\ket{\Uparrow}$ and $\ket{\Downarrow}$, respectively. Note however, that for finite
$N$ the actual ground states in both symmetry sectors are given by
\eq{
|0\rangle_{\mathrm{NS}} = \frac{1}{\sqrt{2}}(\mupr + \mdor), \qquad
|0\rangle_{\mathrm{R}} = \frac{1}{\sqrt{2}}(\mupr - \mdor), \qquad
\label{gsrns}}
which are separated by an exponentially small gap and both have vanishing magnetizations.

\section{Magnetization dynamics\label{sec:mag}}

We are interested in the dynamics of the magnetization of various initial states,
excited locally from the ferromagnetic ground state $\mupr$ and time-evolved
under the Hamiltonian $H$ in \eqref{hxy}. The locality of the excitation is understood
in terms of the Majorana basis, which implies that these excitations may become
highly non-local in the spin-basis representation. In fact, the latter will correspond to
domain-wall excitations and one is interested in how the inhomogeneity spreads
out under unitary time evolution. On the other hand, since the order-parameter
magnetization is not conserved, even a single spin-flip excitation (which is local
in terms of the spins) will lead to nontrivial dynamics.
For the study of domain-wall melting, we will also consider for comparison a local
quench setup where two separate chains are initially prepared in oppositely
magnetized ground states, and subsequently joined together.

The time-evolved magnetization  can be extracted in a number of different ways.
On the numerical side, we apply matrix product state (MPS) calculations
\cite{Schollwoeck11,itensor}
in an open-chain geometry. To ensure that we obtain the proper ferromagnetic
(symmetry-broken) ground state $\mupr$, we introduced a small longitudinal field $h_x > 0$
in the Hamiltonian $H - h_x \sum_i \sigma_i^x$ for the first few sweeps and
set $h_x = 0$ afterwards, until convergence is reached. The excitations are then
created by acting with the matrix product operator representation of the
corresponding spin-excitation. Finally, the time evolution was implemented
with the finite two-site time-dependent variational principle (TDVP) algorithm
\cite{HLOVV14}.

On the other hand, we also employed Pfaffian techniques for the numerical
evaluation of the magnetization. For the simple domain-wall excitation these
were described in Ref. \cite{EME16}, but the calculations can easily be generalized
for the other local excitations we deal with. In all of the examples we observed
a perfect agreement with the results of MPS calculations.

Finally, we also present analytical results based on form-factor calculations.
To this end, one has to first express the excited initial state
$\ket{\psi_0}=(\ket{\psi_0}_\mathrm{R}+\ket{\psi_0}_\mathrm{NS})/\sqrt{2}$
in the fermion basis, which is then time-evolved with the corresponding
Hamiltonian in both symmetry sectors as
\eq{
\ket{\psi_t}_\mathrm{R/NS} = \ee^{-it H_\mathrm{R/NS}}
\ket{\psi_0}_\mathrm{R/NS} \, .
\label{psit}}
Once $\ket{\psi_0}_\mathrm{R/NS}$ is written as a linear combination
of the $K$-particle eigenstates \eqref{hdiag}, the time evolution is trivial
\eq{
\ee^{-it H_\mathrm{R/NS}} |q_1,q_2,\dots,q_K\rangle_{\mathrm{R/NS}} =
\ee^{-it \sum_{k=1}^K \epsilon_{q_k}}|q_1,q_2,\dots,q_K\rangle_{\mathrm{R/NS}} \, ,
\label{qkte}}
since the Hamiltonian $H_\mathrm{R/NS}$ is diagonal in this basis.
It is useful to introduce the normalized magnetization which can
be evaluated as
\eq{
\mnt =
\frac{\rl{\psi_t} \sigma^x_n \nsr{\psi_t}} {\rl{0} \sigma^x_n \nsr{0}} \, .
\label{mnt}}
Note that, since the operator $\sigma^x_n$ changes the parity of the state,
the only non-vanishing contribution to the expectation value is between
different parity sectors. In turn, the calculation of $\mnt$ boils down to
evaluating multiple sums over the momenta with the form factors
$\rl{p_1,\dots,p_L} \sigma^x_n \nsr{q_1,\dots,q_K}$, which are known
explicitly from previous studies \cite{IST11,Iorgov11,IL11}.
In the following we always consider the thermodynamic limit $N \to \infty$,
where the sums over momenta can be turned into integrals and the expressions
for the form factors are summarized in Appendix \ref{app:ff}.

\subsection{Single domain wall}

Our first example is a single domain wall, which has already been considered
for the TI \cite{EME16} as well as for the XY chains \cite{EM18}.
For completeness, we revisit here the results obtained previously for the
normalized magnetization. The single domain wall is an excitation
$\ket{\psi_0} = D_{n_1}\mupr$ 
created by the operator
\eq{
D_{n_1} = \prod_{j=1}^{n_1-1}\sigma^{z}_{j} \, \sigma^{x}_{n_1} = a_{2n_1 -1} \, .
}
As remarked before, $D_{n_1}$ is strictly local in terms of the fermions,
whereas in the spin representation it creates spin-flips all over the sites $j<n_1$.
In the eigenbasis of the Hamiltonian it corresponds to a linear combination
of one-particle states 
\eq{
\ket{\psi_0} = \frac{1}{\sqrt{N}} \sum_{q}
\ee^{-iq(n_1-1/2)} \ee^{i \theta_q/2} \ket{q} ,
}
where we have suppressed the subscripts $\mathrm{R/NS}$ of the symmetry sector
for notational simplicity. One thus only needs the form factors between one-particle
states, which has a relatively simple form \eqref{ff11} given in Appendix \ref{app:ff}.
Performing the time evolution \eqref{psit} via \eqref{qkte} and inserting the result
into \eqref{mnt}, one arrives at
\eq{
\mnt=
\int_{-\pi} ^{\pi} \frac{\dd p}{2\pi} \int_{-\pi}^{\pi} \frac{\dd q}{2\pi}
\frac{\epsilon_p + \epsilon_q}{2\sqrt{\epsilon_p \epsilon_q}}
 \frac{\ee^{i(n-n_1+1/2)(q-p)}}{i \sin \left(\frac{q-p}{2}\right)}
\ee^{i(\theta_q - \theta_p)/2} \ee^{-i(\epsilon_q-\epsilon_p)t} \, .
\label{mntd}}

The above expression simplifies considerably in appropriate scaling limits.
Indeed, noting that the integral receives the dominant contribution due to a
pole at $q=p$ in the integrand of \eqref{mntd}, one can change variables
as $Q=q-p$ and $P=(q+p)/2$, and perform a stationary phase analysis
as described in Appendix \ref{app:sp}. In turn, one obtains
\eq{
\mnt = 
1-2\int_{-\pi}^{\pi} \frac{\dd P}{2\pi}  \Theta(v_P - \nu) \, ,
\qquad \nu =\frac{n-n_1+1/2}{t} \, ,
\label{mntsc}}
which is the so-called hydrodynamic scaling limit. Here $\Theta(x)$ is the
Heaviside step function, $v_P=\frac{\dd \epsilon_P}{\dd P}$ is the group velocity
of the single-particle excitations and $\nu$ is the ray variable, with the distance
measured from the initial location $n_1-1/2$ of the domain wall.
The result \eqref{mntsc} has a simple semiclassical interpretation,
which has been applied many times to understand front dynamics in quantum chains
 \cite{SY97,RI11,DIR11,KMZ17}.
Namely, the magnetization is transported by single-particle excitations, each carrying
an elementary spin-flip, which contribute to the hydrodynamic profile at a given
ray only if their velocity $v_P > \nu$.

Another interesting scaling regime emerges around the edge of the front
$\nu \approx v_{max}$, given by the maximum speed of excitations.
In order to understand the fine structure of the edge, a higher order
stationary phase analysis has to be performed around the momentum
$q_*$ which yields the maximum velocity $v_{q_*}=v_{max}$.
As shown in Appendix \ref{app:sp}, this leads to the following result
\eq{
\mnt \approx 1 - 2 \left(\frac{2}{|v''_{q_*}|t}\right)^{1/3} \rho(X) \, , \qquad
X = (n-n_1+1/2+\theta'_{q_*}/2-v_{q_*}t) \left(\frac{2}{|v''_{q_*}|t}\right)^{1/3}.
\label{medge}}
In other  words, with the proper choice of the scaling variable $X$ measuring
the distance from the edge, and after appropriate rescaling, the fine structure
of the magnetization front is given via the function
\eq{
\rho(X) = \mathcal{K}_{Ai}(X,X)=\left[ \Ai'(X) \right]^2 - X\Ai^2(X) \, .
\label{rhox}}
Note that $\rho(X)$ is nothing else but the diagonal part of the Airy-kernel
$\mathcal{K}_{Ai}(X,Y)$ \cite{TW94},
which appears in a number of front evolution problems related to free-fermion
edge universality \cite{HRS04,ER13,VSDH15,Kormos17,PG17,Fagotti17,BK19,Stephan19}.

The results \eqref{mntsc} and \eqref{medge} have already been tested against
numerical calculations for various parameters of the XY chain, where the notable
feature of a hydrodynamic phase transition at $h_c=1-\gamma^2$ was observed \cite{EM18}.
Indeed, this phase transition can be understood by the appearance of a second local
maximum in the group velocities $v_q$ for $h<h_c$, which in turn leads to kinks
in the bulk of the hydrodynamic magnetization profile \cite{EM18}.

Finally, it should be noted that the analytical result was obtained by following
the time evolution of one-particle states building up the domain wall.
Strictly speaking, these states are eigenstates of $H_-$ only, i.e. the time
evolution has to be performed with antiperiodic boundary conditions on the
spin chain. However, since the form factor calculations are carried out
directly in the thermodynamic limit, the boundaries actually do not play any role.

\subsection{Double domain wall}

We now move on to consider more complicated excitations, that are
created by acting with the operator
\eq{
D_{n_1,n_2} =
\sigma^{x}_{n_1-1}\prod_{j=n_1}^{n_2-1}\sigma^{z}_{j} \, \sigma^{x}_{n_2} =
-i \, a_{2n_1-2} \, a_{2n_2-1} \, ,
\label{dmn}}
where $n_2>n_1$ is assumed.
In terms of fermions this is a two-local operator, i.e. supported on two sites only.
In contrast, $D_{n_1,n_2}$ is again nonlocal in the spin representation,
and it is easy to see that it describes a double domain wall, located
at sites $n_1$ and $n_2$, respectively.
Using \eqref{ab}, the excited initial state can be written as
%
\eq{
\ket{\psi_0} = \frac{1}{N}\sum_q \ee^{iq(n_2-n_1)} \ee^{-i\theta_q} \ket{0} -
 \frac{1}{N}\sum_{q_1,q_2} \ee^{-iq_1(n_1-1/2)}\ee^{-iq_2(n_2-1/2)}
\ee^{-i(\theta_{q_1}-\theta_{q_2})/2} \ket{q_1,q_2} \, .
\label{dd0}}
%
We shall restrict ourselves to the case $n_2-n_1 \gg 1$, i.e. when the
two domain walls are spatially well separated, such that the sum in the
first term of \eqref{dd0} becomes highly oscillatory and can be neglected.
The initial state then involves only two-particle excitations
and the time evolved state can be written as
\eq{
\ket{\psi_t} =
-\frac{1}{N}\sum_{q_1,q_2} \ee^{-iq_1(n_1-1/2)}\ee^{-iq_2(n_2-1/2)}
\ee^{-i(\theta_{q_1}-\theta_{q_2})/2} 
\ee^{-i(\epsilon_{q_1}+\epsilon_{q_2})t}\ket{q_1,q_2} \, .
\label{ddt}}

The magnetization $\mnt$ can thus be expressed as a quadruple integral via 
two-particle form factors $\rl{p_1,p_2} \sigma^x_n \nsr{q_1,q_2},$
that are reported in \eqref{ff22} in Appendix \ref{app:ff}.
The result can be simplified, similarly to the single domain wall case,
by analyzing the pole-structure of the form factors combined with a
stationary phase approximation. The poles appear for momenta
satisfying $q_1=p_1$ and $q_2=p_2$ or $q_1=p_2$ and $q_2=p_1$.
For the first pole one obtains two independent stationary phase conditions
\eq{
v_{P_i}t - (-1)^i \theta'_{P_i} - (n-n_i+1/2) =0 \, ,
\label{spdd}}
where $P_i=(q_i + p_i)/2$ for $i=1,2$.
Note that this pole corresponds to a process where the incoming
momenta are matched with the outgoing ones at each domain wall separately.
In contrast, at the second pole an incoming momentum of the first domain wall
must match with an outgoing momentum of the second domain wall.
However, as shown in Appendix \ref{app:sp}, after the exchange of the
outgoing momenta and under the assumption $n_2-n_1 \gg 1$,
the stationary phase condition cannot be satisfied.
Thus only the first pole gives a contribution to the integral 
 and leads to the result
\eq{
\mathcal{M}_n(t) = \prod_i \int \frac{\dd P_i}{2\pi}
\big[1-2 \, \Theta \left(v_{P_i}-\nu_i\right) \big] ,\qquad
\nu_i = \frac{n-n_i+1/2}{t} \, .
\label{mntdd}}

The hydrodynamic scaling limit of the profile in \eqref{mntdd} has thus a factorized
form with again a very simple physical interpretation.
The ray variables $\nu_i$ now measure the distances from the corresponding
initial domain wall locations $n_i-1/2$, where quasiparticles with velocity $v_{P_i}$
are emitted, each carrying a spin-flip. If, for a given pair of particles, one has
$v_{P_1}>\nu_1$ and $v_{P_2}>\nu_2$ then both of the particles have reached
site $n$ at time $t$, hence the spin is flipped twice and one has a positive
contribution. If, on the other hand, $v_{P_1}<\nu_1$ and $v_{P_2}>\nu_2$,
then only one particle has arrived and the contribution is negative.
The profile is then obtained by summing the contributions over all pairs.

%
\begin{figure}[htb]
\center
\includegraphics[width=0.49\columnwidth]{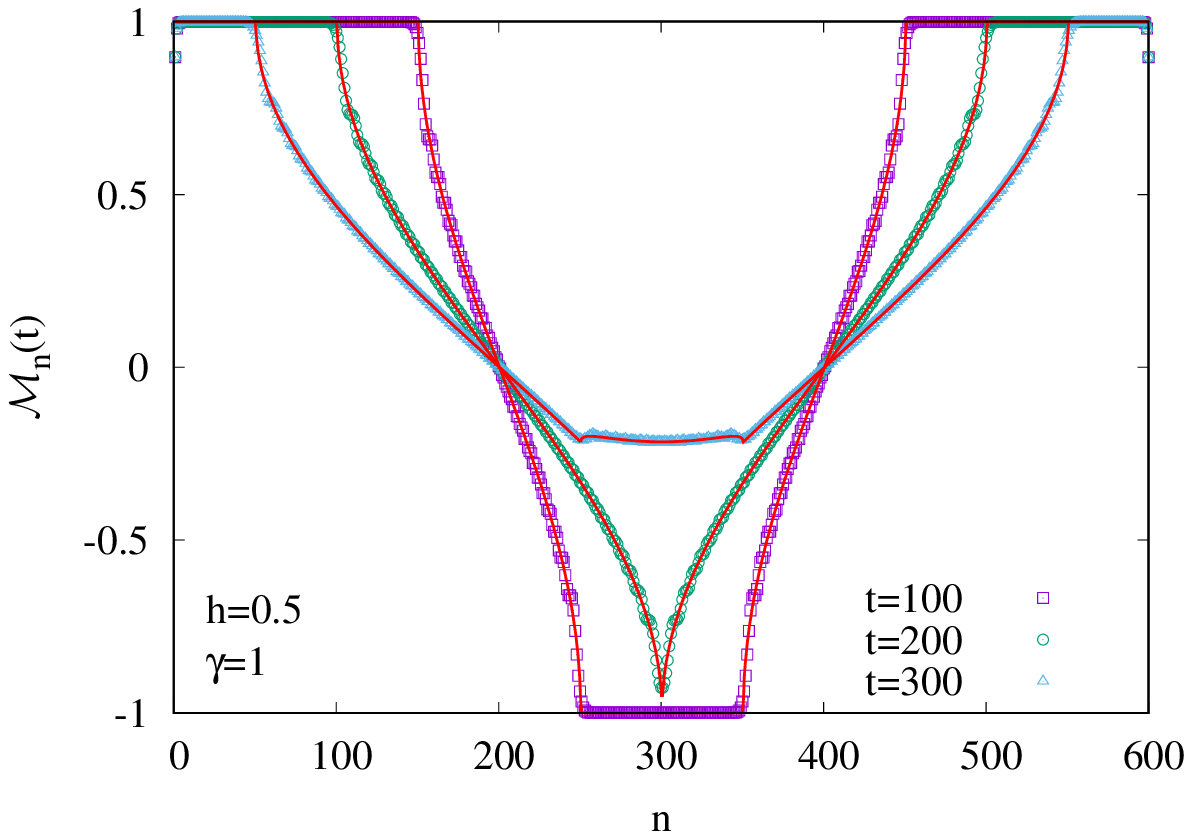}
\includegraphics[width=0.49\columnwidth]{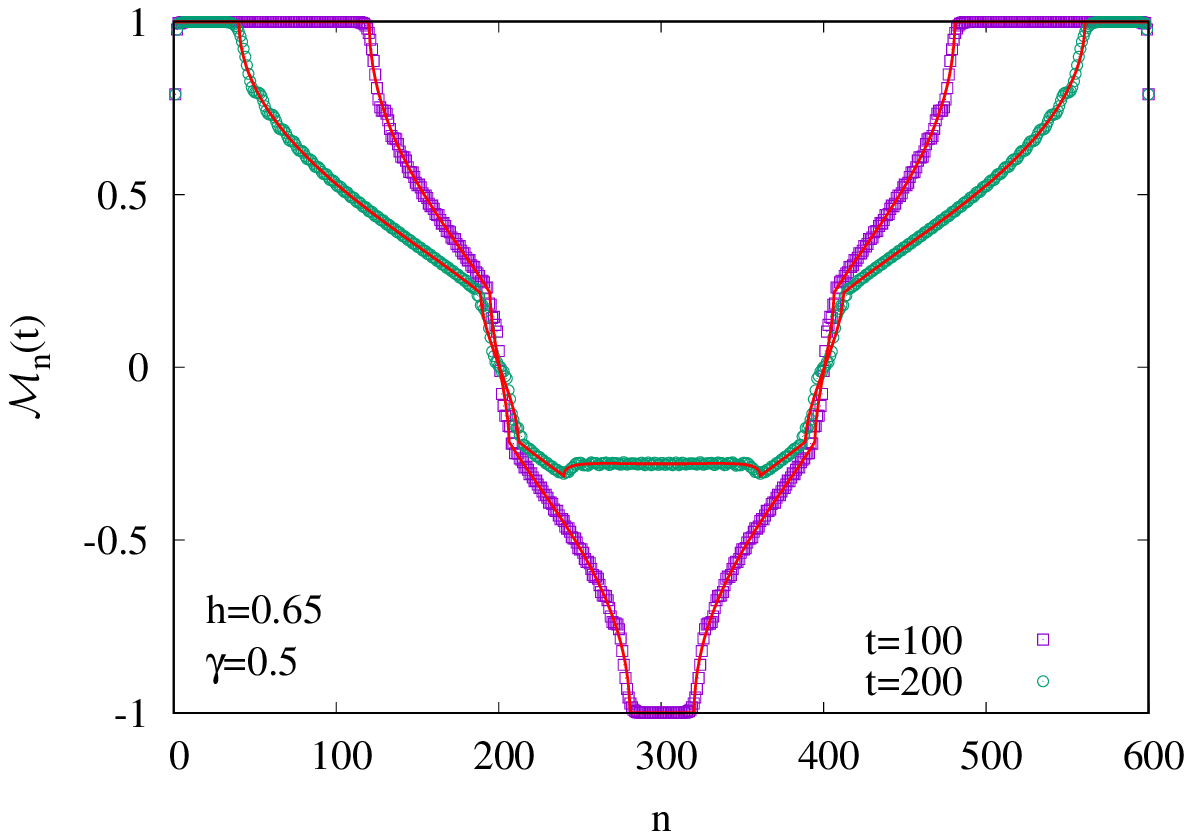}
\caption{Magnetization profiles after a double domain wall excitation for different
times and various $h$ and $\gamma$.
The solid red lines show the approximation \eqref{mntdd}.
The parameters are $N=600$, $n_1=201$ and $n_2=401$.}
\label{fig:mntdd}
\end{figure}
%

In Fig. \ref{fig:mntdd} we show the results of our MPS simulations together
with the result \eqref{mntdd}. One can see a perfect agreement, even after the
two fronts propagating from different locations overlap in the middle.
In particular, one observes the emergence of two cusps at the ends of the
overlap region, which follows from the factorized form of \eqref{mntdd},
i.e. one multiplies two single domain wall front profiles, each having 
square-root singularities at their edges. Moreover, this also implies
that the outer edge of the front is still described by the same scaling \eqref{medge}
as for the single domain wall. On the right of Fig. \ref{fig:mntdd} there are
extra kinks to be seen, which is due to the fact that one has $h<h_c$ there,
i.e. one is beyond the hydrodynamical phase transition point.

\subsection{Single spin-flip\label{sec:magsz}}

After having discussed the evolution of domain walls, we now study
a very simple excitation, in the form of a single flipped spin.
Naively, one would think that this excitation has a trivial hydrodynamic limit,
and the flipped spin just disperses. However, since the magnetization is not
conserved under the XY dynamics, it turns out that the profile is far from
being trivial. In fact, the operator that creates a spin-flip at site $n_1$ is just 
$\sigma^z_{n_1} = -i a_{2n_1-1}a_{2n_1}$, which is strictly local in the
spin representation, but is again two-local, i.e. a product of two adjacent
Majoranas in the fermionic picture.
Hence, this form is more reminiscent of a double domain wall excitation, with
the exception that they are now created at neighbouring sites.
Rewriting the excitation in the fermionic basis one has
\eq{
\ket{\psi_0} = 
m^z \ket{0} -
\frac{1}{N}\sum_{q_1,q_2} \ee^{-iq_1(n_1-1/2)}\ee^{-iq_2(n_1+1/2)}
\ee^{i(\theta_{q_1}-\theta_{q_2})/2} \ket{q_1,q_2} \, ,
\label{sz0}}
where the ground-state contribution is now proportional to the transverse
magnetization
\eq{
m^z = \langle 0 |\sigma^z_n|0\rangle =
-\int_{-\pi}^{\pi}\frac{\dd q}{2\pi} \ee^{i(\theta_q+q)} \, ,
\label{mz}}
and thus cannot be neglected.

The calculation of $\mnt$ follows the same steps as in the previous cases.
Note, in particular, that the two-particle contribution in \eqref{sz0} has almost the
same form as \eqref{dd0} for the double domain wall with $n_2=n_1+1$,
except for the sign of the Bogoliubov phases. After time evolving and taking the
expectation value with $\ket{\psi_t}$, one has now cross terms where the form
factors $\rl{0} \sigma^x_n \nsr{q_1,q_2}$ appear, see \eqref{ff20}. However,
since they have no poles, it is easy to see that their contribution is negligible
in the scaling limit we are interested in. On the other hand, the two-particle form
factors now yield a contribution from both of the poles. Indeed, the stationarity
condition is, up to the sign of the $\theta'_{P_i}$ term, is the same as \eqref{spdd}
for the double domain wall with $n_2=n_1+1$. However, in the limit of $t \gg 1$
and $|n-n_1| \gg 1$, the two equations are essentially the same.
Hence, the process in which an incoming momentum of the first domain wall
scatters into an outgoing momentum of the neighbouring one
is equally well permitted and yields a sizable contribution.

\begin{figure}[htb]
\center
\includegraphics[width=0.49\columnwidth]{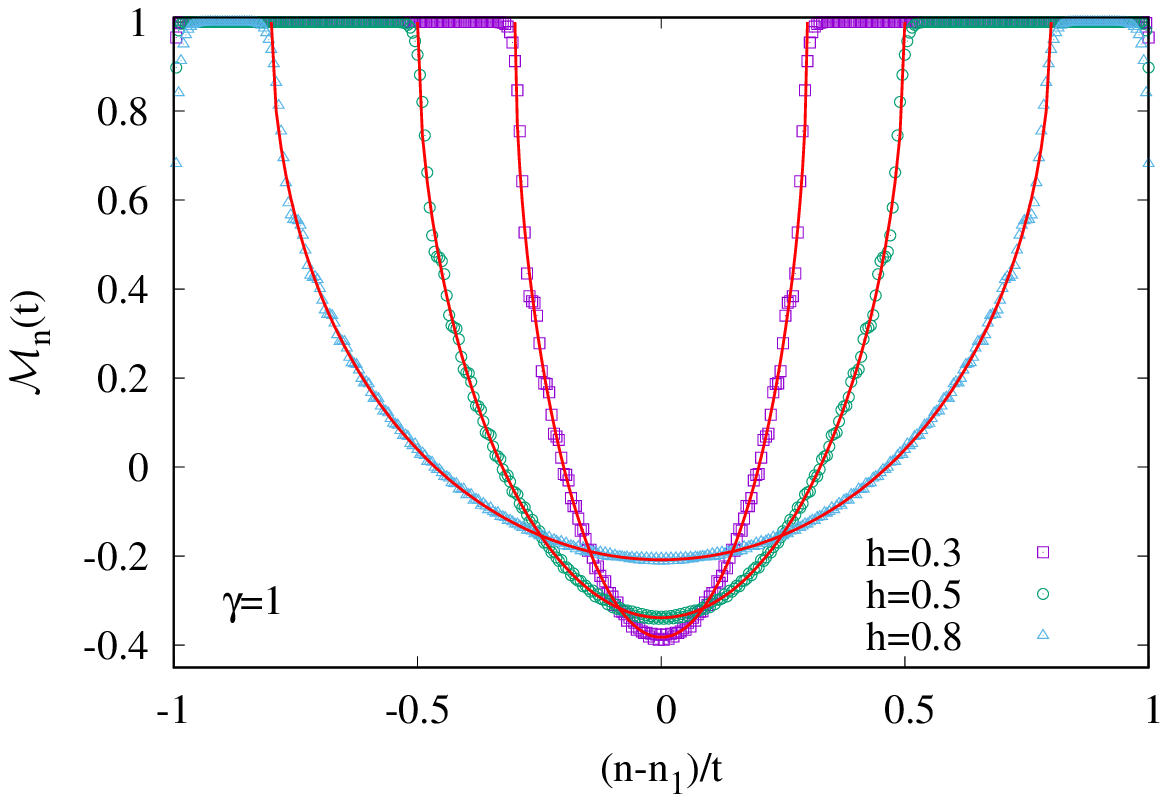}
\includegraphics[width=0.49\columnwidth]{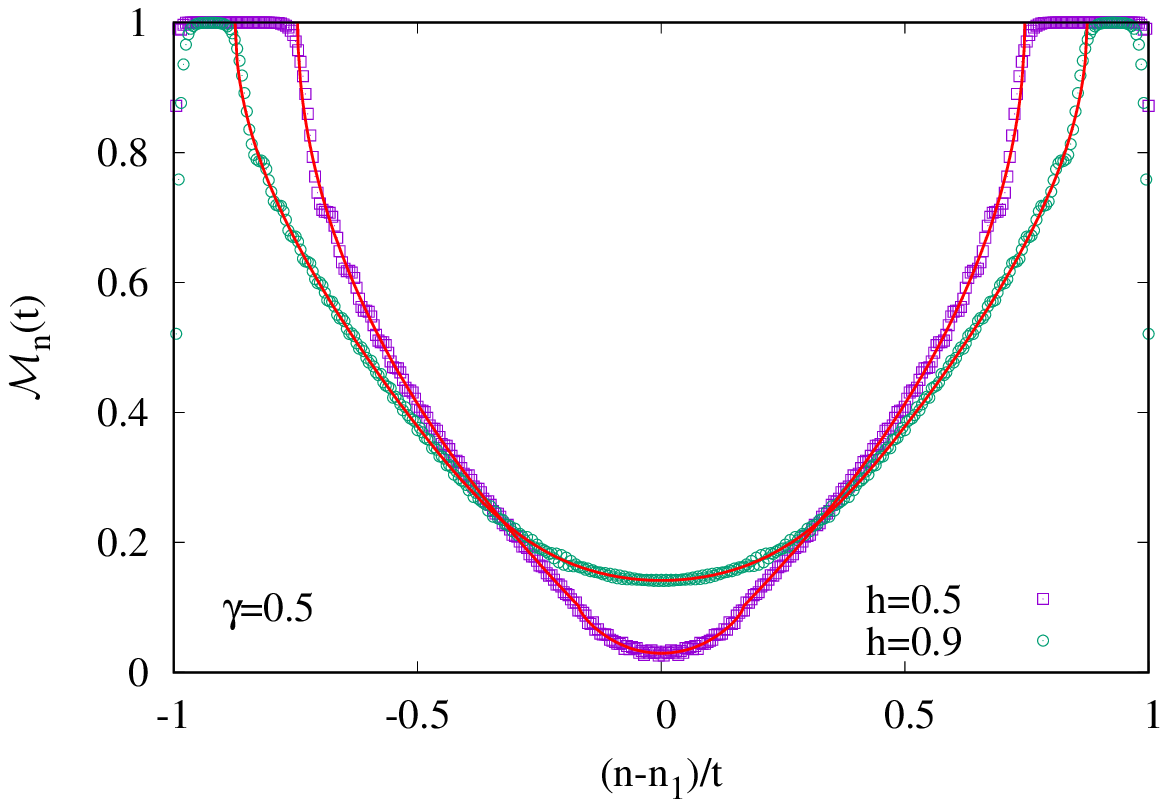}
\caption{Magnetization profiles after a spin-flip excitation for various $h$ and $\gamma$.
The red solid lines show the approximation \eqref{mntsz}.
The parameters are $N=400$, $n_1=200$ and $t=200$.}
\label{fig:mntsz}
\end{figure}

Carrying out the stationary phase analysis in detail (see Appendix \ref{app:sp}),
one arrives at the following result in the hydrodynamic limit
\eq{
\mathcal{M}_n(t) = (m^z)^2
+\left[1-2 \int_{-\pi}^{\pi} \frac{\dd P}{2\pi} \, \Theta \left(v_P - \tilde\nu \right) \right]^2
- \left|m^z+2 \int_{-\pi}^{\pi} \frac{\dd P}{2\pi} \ee^{iP}\ee^{i\theta_{P}}
\Theta \left(v_P - \tilde\nu \right) \right|^2 ,
\label{mntsz}}
where the ray variable $\tilde \nu=\frac{n-n_1}{t}$ is slightly changed compared to \eqref{mntsc},
since the distance is now measured from the location $n_1$ of the spin-flip.
The profile can be written as the sum of three terms, where the first one is simply the
ground-state contribution. The second one corresponds to the factorized result for the
double domain wall and the third one describes a kind of interference term,
where the momenta of the excitations building up the two domain walls are exchanged.
There is no simple semiclassical interpretation of this interference term,
since the quasiparticles contribute with a phase factor. 
The result \eqref{mntsz} is compared against our numerical calculations in
Fig. \ref{fig:mntsz} with an excellent agreement.

%
%

\begin{figure}[htb]
\center
\includegraphics[width=0.49\columnwidth]{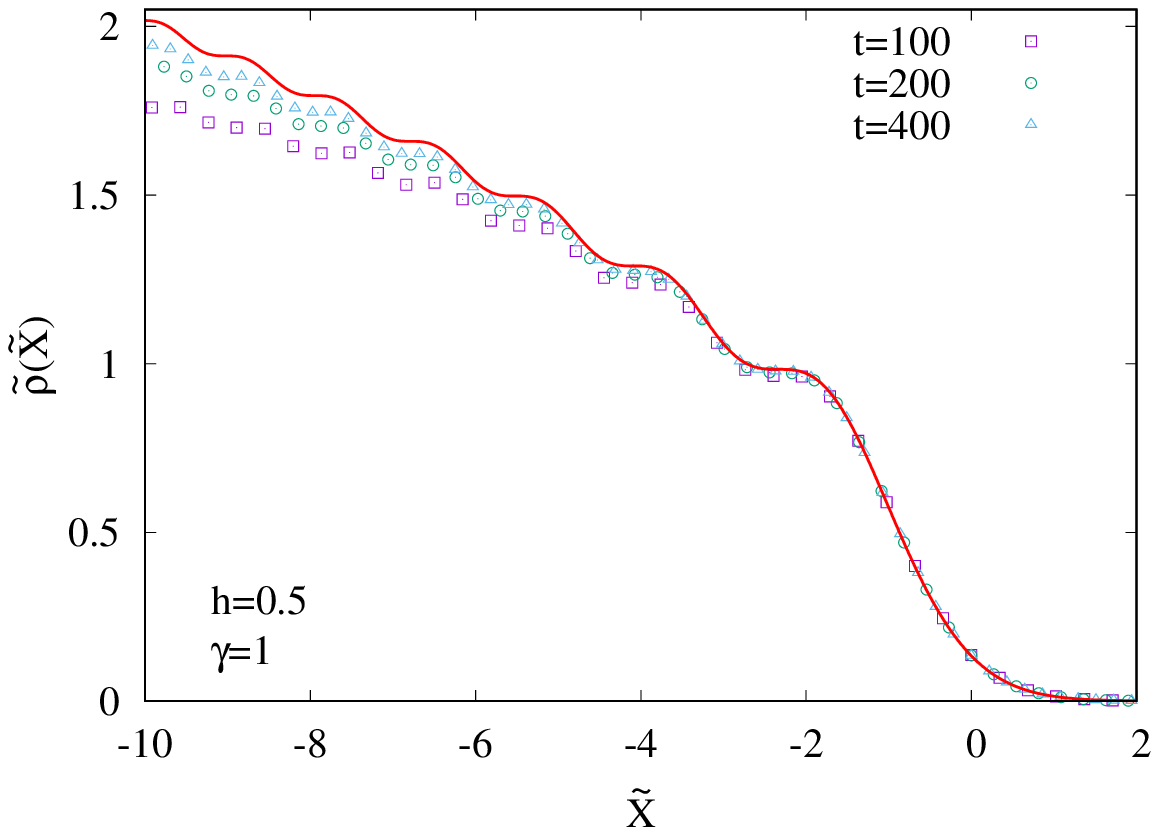}
\includegraphics[width=0.49\columnwidth]{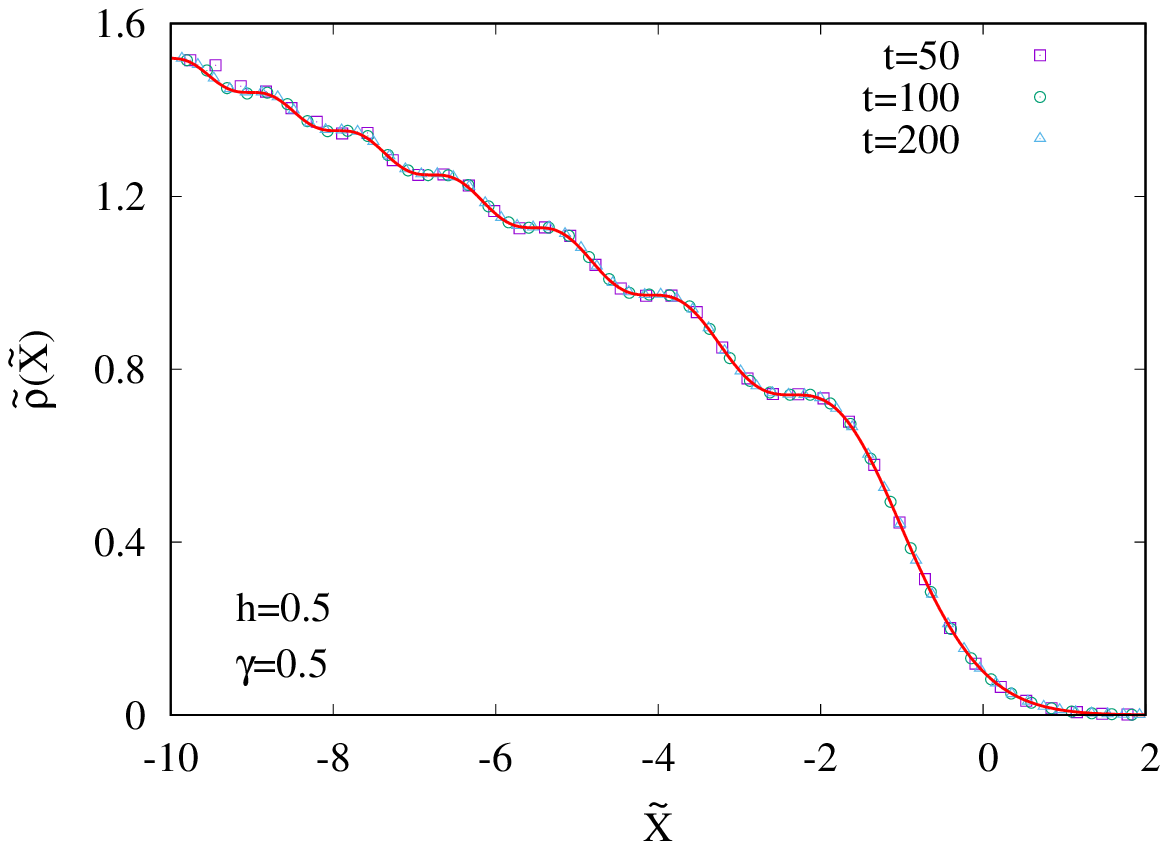}
\caption{Edge scaling \eqref{mszedge} of the magnetization profiles after a spin-flip
excitation for various $h$ and $\gamma$. The red solid lines show the scaling function in
\eqref{rhot}.}
\label{fig:mszedge}
\end{figure}

It is also interesting to have a look at the edge behaviour of the profile.
Performing the higher order stationary phase analysis (see Appendix \ref{app:sp}),
one is led to the following result
\eq{
\mnt \approx 1 - 2 \left(\frac{2}{|v''_{q_*}|t}\right)^{1/3} \tilde \rho(\tilde X) \, , \qquad
\tilde X = (n-n_1-v_{q_*}t) \left(\frac{2}{|v''_{q_*}|t}\right)^{1/3} ,
\label{mszedge}}
where the scaling function is given by
\eq{
\tilde \rho(\tilde X) = \big[ 2 + 2 \, m^z \cos(\theta_{q_*}+q_*) \big]
\mathcal{K}_{Ai}(\tilde X, \tilde X) \, .
\label{rhot}}
The result is thus very similar to the one for the domain wall in \eqref{medge},
however the scaling function $\tilde \rho(\tilde X)$ acquires a nontrivial prefactor,
which depends explicitly on the transverse magnetization $m^z$, and even
on the Bogoliubov phase evaluated at $q_*$ where the quasiparticle velocity
has its maximum. In particular, this phase factor vanishes for the TI chain
and one has a factor of $2$ difference with respect to $\rho(X)$. This explains
the numerical findings of Ref. \cite{ZGEN15} where the very same setup was studied.
We checked the validity of the edge scaling \eqref{mszedge} in Fig. \ref{fig:mszedge}
for various parameter values and found a very good agreement,
there are however some differences in the convergence towards
the scaling function $\tilde \rho(\tilde X)$.

%
%

\subsection{Local quench}

As a final example, we show here the results for the magnetization profile
resulting from a local quench. That is, instead of applying a local excitation
to the symmetry-broken ferromagnetic state, we rather prepare the two halves
of our chain in oppositely magnetized ground states and join them together.
Our goal is to check whether this protocol yields a similar result for the
hydrodynamic profile as the one found for the single domain wall excitation. 

\begin{figure}[htb]
\center
\includegraphics[width=0.49\columnwidth]{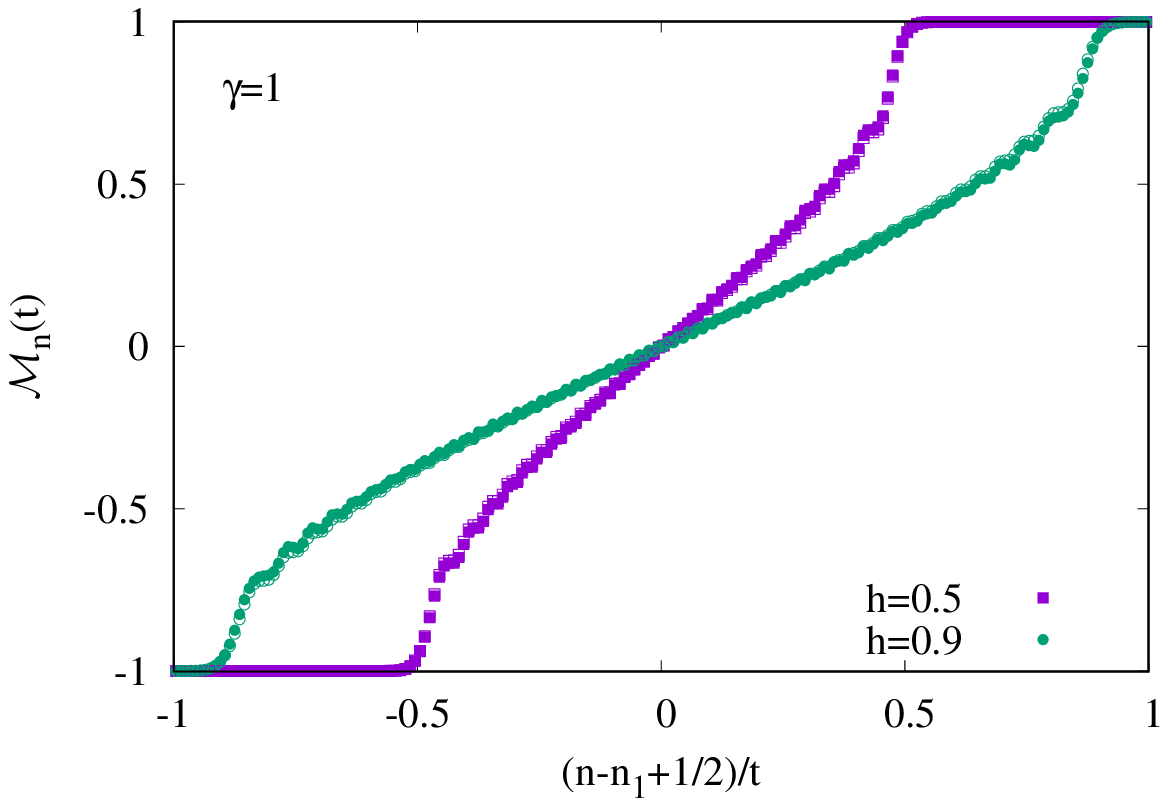}
\includegraphics[width=0.49\columnwidth]{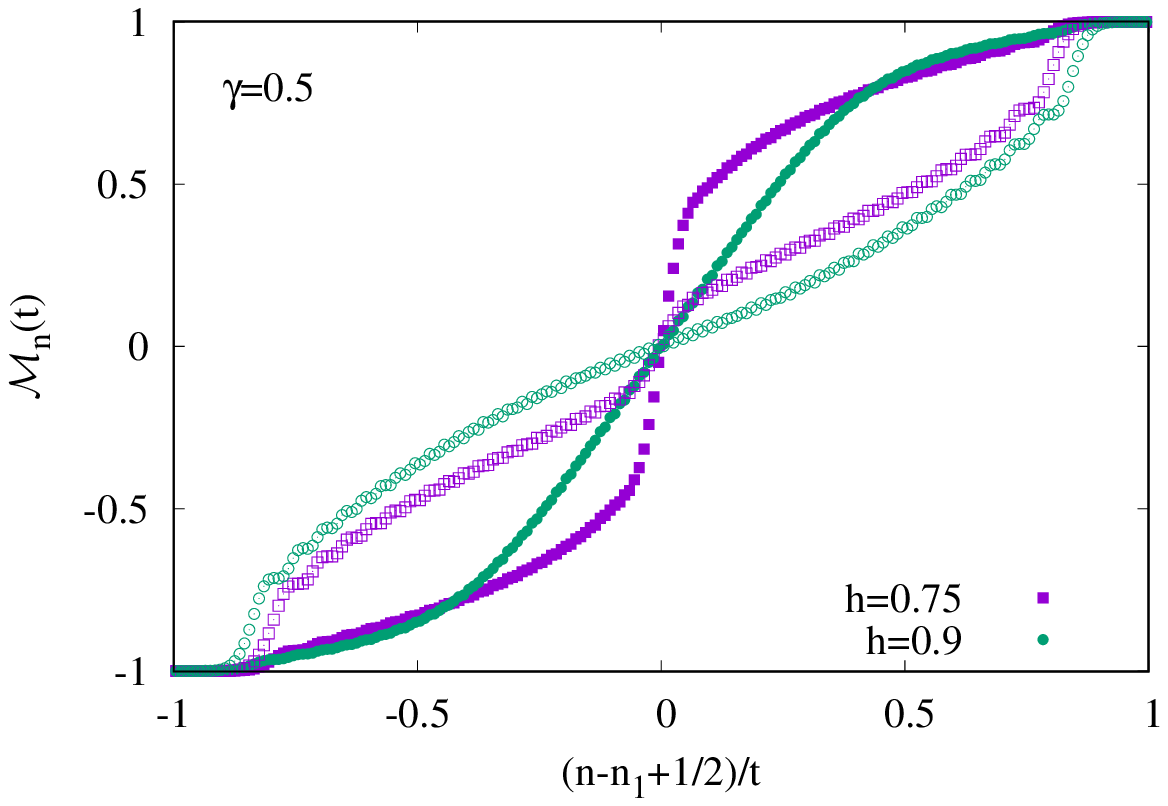}
\caption{Magnetization profiles after the local quench (full symbols) vs.
single domain wall excitation (empty symbols), for various $h$ and $\gamma$.
The parameters are $N=400$, $n_1=201$ and $t=100$.}
\label{fig:mlq}
\end{figure}

The initial and time-evolved states are now given by
\eq{
\ket{\psi_0} = \ket{\Downarrow} \otimes \ket{\Uparrow}, \qquad
\ket{\psi_t} = \ee^{-iHt}\ket{\psi_0} .
}
Since our initial state is not prepared as an excitation upon the bulk vacuum
state, it is a nontrivial question how $\ket{\psi_0}$ can be written in the basis
of the full Hamiltonian $H$. Thus we shall only perform numerical
(MPS and Pfaffian based) calculations for the quench.
The results, shown in Fig. \ref{fig:mlq}, turn out to be rather surprising.
Namely, we find that in the TI limit ($\gamma=1$) the profiles after the local
quench (full symbols) almost exactly coincide with the ones for the domain
wall excitation (empty symbols). The only deviations visible at the scale of the
figure are around the front edges. In sharp contrast, for $\gamma=0.5$
one has a huge deviation between the profiles for all the values of $h$
we considered. This signals that in the latter case the factorized
initial state is not well approximated by a single-particle excitation
in the fermionic basis. We observe that the mismatch between the profiles
gradually increases as one moves away from the TI limit. However,
we have no clear explanation of this phenomenon which needs further studies.

%
%

\section{Correlation functions\label{sec:corr}}

The form-factor approach is not restricted to the study of the magnetization
profile. The next simplest physically interesting observable is the correlation
function between the spins. Here we shall restrict ourselves to equal-time
correlations between the $x$-components of the spin, which have already
been addressed briefly in \cite{EM18}. It is useful to work with the normalized
correlation functions
\eq{
\cmnt= \nsl{\psi_t} \hat{\mathcal{M}}_m \hat{\mathcal{M}}_n \nsr{\psi_t} \, ,
\label{cmn}}
where the expectation value is now taken between the NS components only,
since the operator $\sigma^x_m\sigma^x_n$ does not change the parity.
Note that we use here that the corresponding expectation value between
the R components is equal to \eqref{cmn} in the thermodynamic limit.

In order to get a form-factor expansion of \eqref{cmn}, we shall insert
the resolution of the identity
\eq{
\identity = 
\ket{0}\bra{0} +
\sum_{p} |p\rangle \langle p | +
\sum_{p_1,p_2} |p_1,p_2\rangle \langle p_1,p_2 | +
\sum_{p_1,p_2,p_3} |p_1,p_2,p_3\rangle \langle p_1,p_2,p_3 | + \dots
\label{resid}}
Note that the resolution must be taken within the R sector, but we
omit here the subscripts for notational simplicity.
The form-factor expansion can be obtained by inserting the expression of
$\nsr{\psi_t}$ in terms of the fermionic basis. We focus here on the case
of a single domain wall, since the calculations become rather cumbersome
for more complicated excitations. In this case $\nsr{\psi_t}$ is a
superposition of single-particle states only and it is reasonable to assume
that, for distances much larger than the correlation length $|n-m|\gg \xi$,
the dominant contribution to the correlations comes from the single-particle
terms in \eqref{resid} as well. To lowest order in the form-factor expansion
we thus arrive at the result
%
\begin{align}
\cmnt \simeq
&\int \frac{\dd q_1}{2\pi}\int \frac{\dd q_2}{2\pi}
\ee^{-i(\theta_{q_1}-\theta_{q_2})/2}\ee^{i(\epsilon_{q_1} -\epsilon_{q_2})t}
\nonumber \\  \times
&\int \frac{\dd p}{2\pi} \frac{\epsilon_p + \epsilon_{q_1}}{2\sqrt{\epsilon_p \epsilon_{q_1}}}
\frac{\epsilon_p + \epsilon_{q_2}}{2\sqrt{\epsilon_p \epsilon_{q_2}}}
\frac{\ee^{-i(m-n_1+1/2)(q_1-p)}}{\sin \frac{q_1-p}{2}}\frac{\ee^{i(n-n_1+1/2)(q_2-p)}}{\sin \frac{q_2-p}{2}} \, .
\label{cmnint}
\end{align}

The hydrodynamic limit of \eqref{cmnint} can be obtained in a similar fashion as was
done for the magnetization profile. Expanding around the poles of the integrand
and using the properties of the $\Theta$ function (see Appendix \ref{app:cf} for details)
one obtains
\eq{
\cmnt \simeq
1-2\int_{-\pi}^{\pi} \frac{\dd P}{2\pi}  \Theta(v_{P} - \mu) \Theta(\nu - v_{P}) \, ,
\label{cmnsc}}
where the ray variables
\eq{
\mu = \frac{m-n_1+1/2}{t} \, , \qquad
\nu = \frac{n-n_1+1/2}{t} \,
}
are measured from the initial domain wall location and the expression has a
very simple interpretation. Let us assume $\nu > \mu$
and consider the contribution of a single quasi-particle traveling at speed $v_P$.
Now, for short times $v_P<\mu$ the excitation has not yet reached the first spin
and thus the correlations are ferromagnetic. Once $\mu < v_P < \nu$, the first spin
has been flipped while the second one is still untouched, hence the correlation
is antiferromagnetic. Finally, after the excitation has traveled through, $v_P > \nu$,
the second spin is also flipped and the correlation becomes ferromagnetic again.

It turns out that, instead of approximating the integrals in \eqref{cmnint}, there
is a way to directly relate $\cmnt$ to the profile $\mnt$. Indeed, by turning the
integral over $p$ into a contour integral and applying the residue theorem,
one obtains the formula \eqref{cvsm1p} reported in Appendix \ref{app:cf},
which is an exact relation at the level of one-particle form factors.
However it is easy to see that, similarly to the hydrodynamic approximation
in \eqref{cmnsc}, it yields perfect ferromagnetic correlations $\cmnt \simeq 1$
when both spins are outside the front region.
Indeed, it can be shown that the many-particle form factors are the ones
responsible for the exponentially decaying correlations $\mathcal{C}^0_{m,n}$
in the ground state \cite{Iorgov11}. One can thus reincorporate these correlations
into the approximation as
\eq{
\cmnt \simeq \mathcal{C}^0_{m,n} + \mathcal{M}_{m}(t)-\mathcal{M}_{n}(t) \, .
\label{cvsm}}

\begin{figure}[htb]
\center
\includegraphics[width=0.6\columnwidth]{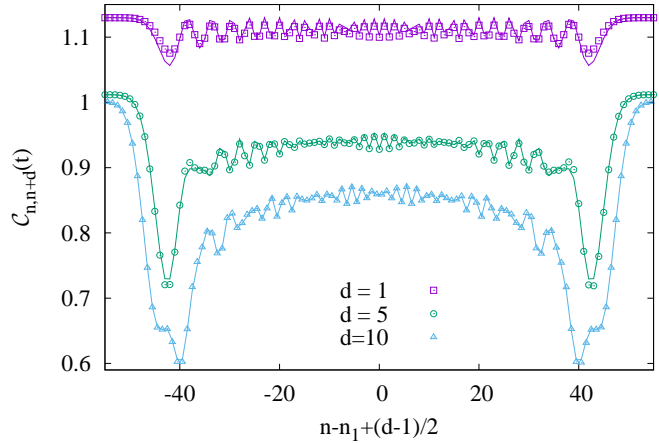}
\caption{Equal-time correlation functions at $t=50$ for the TI model at $h=0.9$,
for various distances $d$ between the spins.
The solid lines show the approximation in \eqref{cvsm}.}
\label{fig:cmnt}
\end{figure}

The relation in \eqref{cvsm} is tested against exact numerical calculations
for the TI chain in Fig. \ref{fig:cmnt}. We have calculated the correlations
along the front region while keeping the distance $d$ between the spins fixed.
One can see that, for $d=1$, there is still a slight deviation from \eqref{cvsm}
which, however, decreases with increasing $d$. For $d=10$ one has already
an excellent agreement with no visible deviations. In fact, for $|n-m| \gg \xi $
one has $\mathcal{C}^0_{m,n}\to 1$, and one recovers the one-particle result
 \eqref{cvsm1p} which should become exact. Note, however, that calculating the corrections
to $\eqref{cvsm}$ would require to evaluate multiple integrals with higher-order
offdiagonal form factors and is thus a difficult task. Nevertheless, a closer investigation
of the form-factor structure in \eqref{ffti} confirms, that the dominant pole contribution
is suppressed and thus one indeed obtains subleading terms.

%
%

\section{Entanglement dynamics\label{sec:ent}}

So far we have studied the simplest observables.
One can, however, gather important information about the time-evolved state
by looking at the entanglement dynamics. In particular, we are interested
in the entanglement profiles along the front region, considering
a bipartition into two disjoint segments $A=\left[1,N/2+r\right]$ and
its complement $B$, and calculating the resulting von Neumann entropy.
Entanglement profiles for domain-wall type initial conditions have been studied 
extensively for time evolution under critical Hamiltonians
\cite{GKSS05,EIP09,SM13,EP14,AHM14,ZGEN15,EB17,GE19,MPP19},
and even a description in terms of CFT has been given \cite{ADSV16,DSVC17}.
However, much less is known about the non-critical case, such as the one at hand.

The calculation of the entanglement profile is straightforward in MPS calculations,
however, extracting the entropy via covariance-matrix techniques for Gaussian states
\cite{Vidal03,PE09} requires some extra considerations. Indeed, the problem
lies in the nature of the initial state, since the excitations are created upon
the symmetry-broken ground state, which is inherently non-Gaussian \cite{FE13}.
Nevertheless, this difficulty can be overcome by relating the problem to
the one where the very same excitations are created upon the Gaussian,
non-magnetized ground states in \eqref{gsrns}.
The method has already been outlined in \cite{EM18} but we expand
here the arguments for completeness.

Let us consider initial states corresponding to the two symmetry-broken
ground states of the system. Using \eqref{gsrns}, the density matrices are given by
\eq{
\mupr \mupl = \rho_e + \rho_o \, , \qquad
\mdor \mdol = \rho_e - \rho_o \, , \qquad
\label{updown}}
where the even and odd parity components, satisfying
$\left[\mathcal{P},\rho_e\right]=0$ and $\left\{\mathcal{P},\rho_o\right\}=0$,
respectively, are defined as
\eq{
\rho_e = \frac{1}{2} \big( \nsr{0} \, \nsl{0} + \rr{0} \, \rl{0} \big) \, , 
\qquad
\rho_o = \frac{1}{2}(\nsr{0} \, \rl{0} + \rr{0} \, \nsl{0}) \, .
\label{rhoeo}}
Clearly, the problem is with the odd component $\rho_o$, since a Gaussian
density operator is by definition even. One can, however, eliminate $\rho_o$
by considering an equal-weight convex combination of the density matrices
in \eqref{updown}. The resulting density matrix $\rho_e$ is itself still a convex
combination of two Gaussian states from the NS and R sectors.
However, working in the thermodynamic limit, these two states become
indistinguishable \cite{FE13}, and one concludes that $\rho_e$ is equivalent
to a proper Gaussian state.

Furthermore, as shown in Ref. \cite{CR17}, excitations that can be written
as a product of Majorana fermions
\eq{
D_J = \prod_{j \in J} a_{j} \, ,
\label{dj}}
where $J$ is an arbitrary index set, preserve Gaussianity.
So does unitary time evolution governed by a quadratic Hamiltonian.
Hence, introducing the notation 
\eq{
\rhoua = \trb \left[ \ee^{-iHt} D^{\phantom{\dag}}_J \mupr \mupl D^\dag_J \, \ee^{iHt}\right] ,
\qquad
\rhoda = \trb \left[ \ee^{-iHt} D^{\phantom{\dag}}_J \mdor \mdol D^\dag_J \, \ee^{iHt}\right] ,
}
for the \emph{reduced} density matrices of a given bipartition, after exciting
and time evolving the initial states in \eqref{updown}, we finally come to the
conclusion that
\eq{
\rho_A = \frac{\rhoua + \rhoda}{2}
\label{rhog}}
is a well-defined Gaussian state living on the Hilbert space of segment $A$.

Our goal is now to relate the entropy $S(\rhoua)=- \Tr \rhoua \ln \rhoua$
of our target state to that $S(\rho_A)$ of the Gaussian state in \eqref{rhog}.
To this end we use the inequality for convex combinations of
density matrices \cite{LR68,Wehrl78} 
\eq{
S \big(\sum_i \lambda_i \rho_i \big) \le
\sum_i \lambda_i S(\rho_i) - \sum_i \lambda_i \ln \lambda_i \, .
\label{sineq}}
First, we note that from trivial symmetry arguments one has 
$S(\rhoda)=S(\rhoua)$. Furthermore, it is also known
\cite{Wehrl78} that the inequality \eqref{sineq} is saturated if the
ranges of $\rho_i$ are pairwise orthogonal, which is again clearly satisfied
in our case due to $\braket{\Uparrow | \Downarrow}=0$.
Hence one finds
\eq{
S(\rhoua) = S(\rho_A) - \ln 2 \, .
\label{sup}}

Finally, it remains to calculate the covariance matrix $\Gamma_A$
corresponding to $\rho_A$, from which the calculation of the entropy
$S(\rho_A)$ follows standard procedure \cite{Vidal03,PE09}.
Since $\rho_A$ is the reduced density matrix of the
time-evolved and excited ground state $\nsr{\psi_t}$,
$\Gamma_A$ is just the reduction of the full covariance
matrix with elements $\Gamma_{k,l}= \nsl{\psi_t}\left[a_k,a_l\right]\nsr{\psi_t}/2$.
This can be obtained by working in the Heisenberg picture.
Since $D_J$ is unitary, $D_J^{\phantom{\dag}}D_J^\dag = \identity$,
the effect of the excitation can be absorbed by a change of the Majorana basis \cite{CR17}
\eq{
a'_k =  D_J^\dag \, a_k \, D_J^{\phantom{\dag}}=
\sum_{l=1}^{2N} Q_{k,l} a_l \, .
}
The orthogonal transformation $Q$ has a simple diagonal matrix form
\eq{
Q_{kl} = \delta_{k,l} \prod_{j\in J} (2\delta_{k,j}-1) \, ,
}
with entries $\pm 1$, depending on whether the corresponding column
is part of the index set $J$ or not. In complete analogy, the unitary time
evolution corresponds to the basis rotation
\eq{
a'_k(t) = \ee^{iHt} a'_k \ee^{-iHt}=\sum_{l=1}^{2N} R_{k,l} a'_l \, ,
}
where the explicit form of the orthogonal matrix $R$ was reported in
Ref. \cite{EME16}. Putting everything together, one finds that
\eq{
\Gamma = R \, Q \, \Gamma_0 \, Q^T R^T ,
\label{tg}}
where $\Gamma_0$ is the ground-state covariance matrix with
elements $(\Gamma_{0})_{k,l}= \nsl{0}\left[a_k,a_l\right]\nsr{0}/2$.

We are now ready to discuss the entanglement dynamics for the
simple excitations introduced in Sec. \ref{sec:mag}.
In each case we have verified that the entropy obtained by the procedure
outlined above agrees perfectly with the results of our MPS calculations.

\subsection{Single domain wall}

\begin{figure}[htb]
\center
\includegraphics[width=0.49\columnwidth]{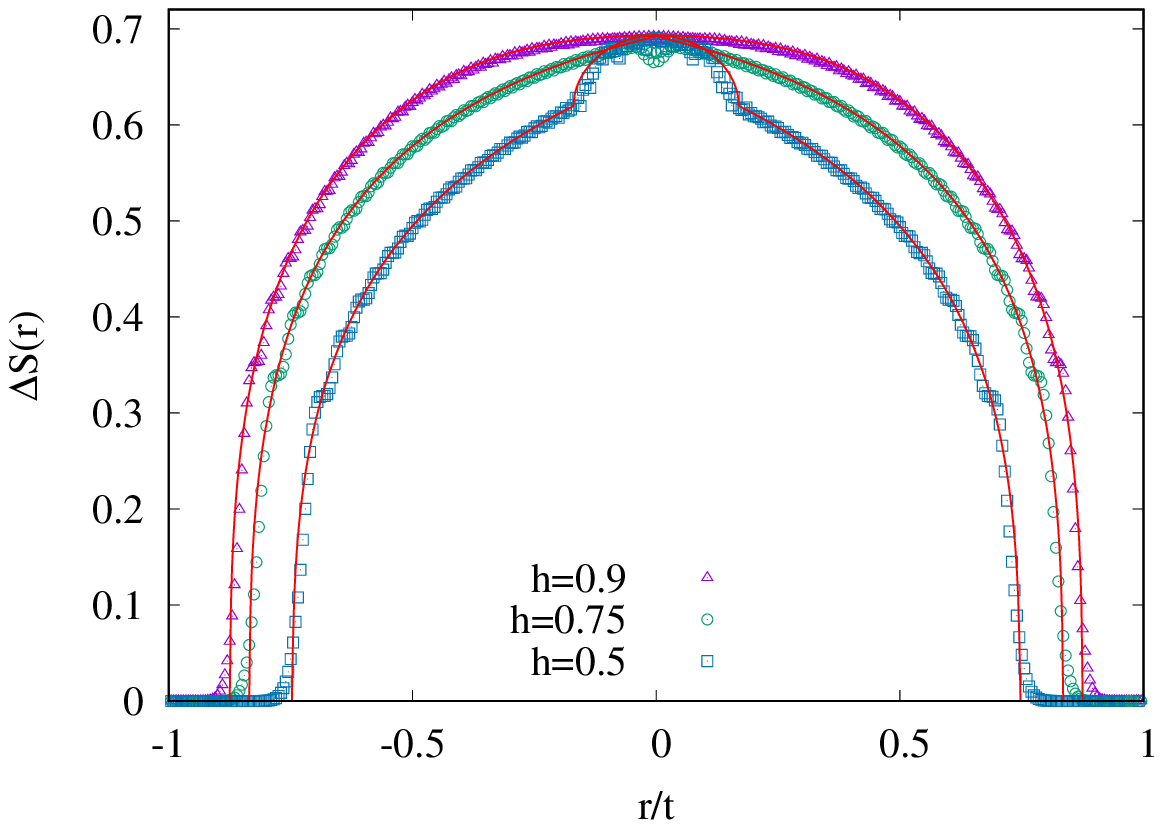}
\includegraphics[width=0.49\columnwidth]{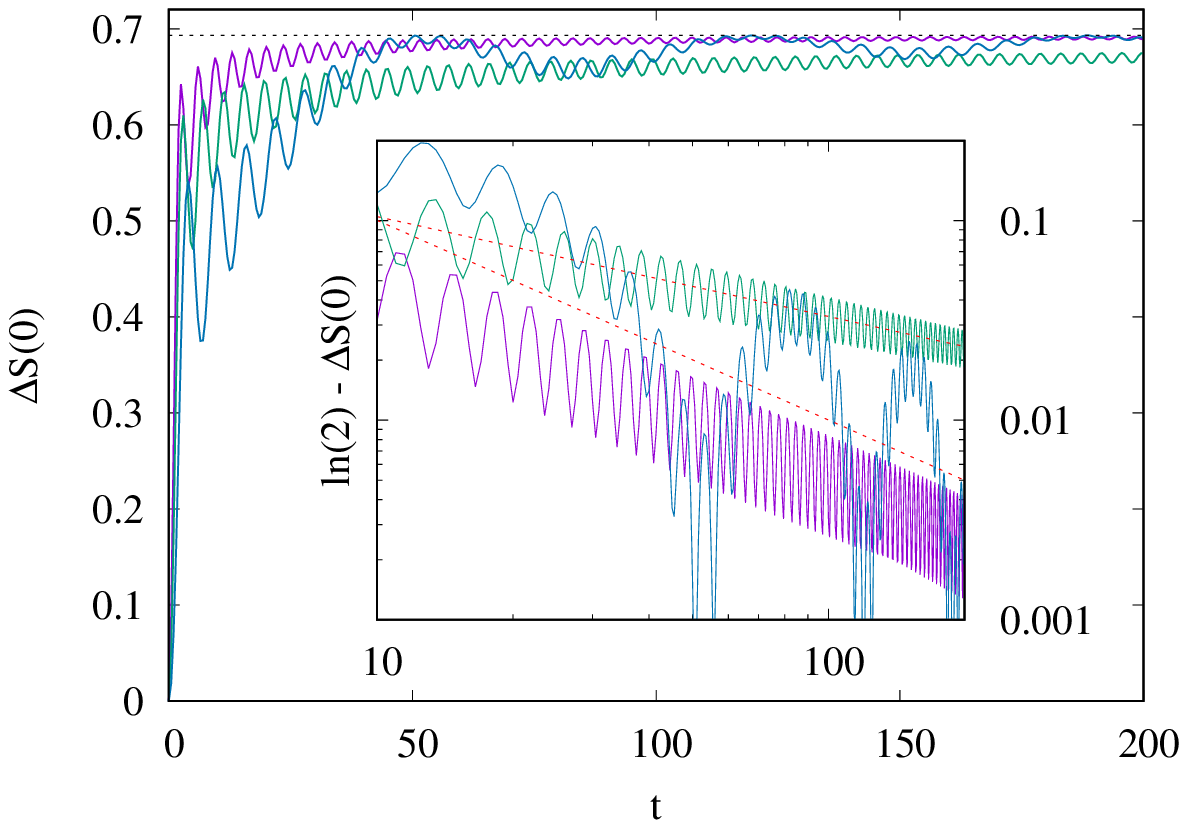}
\caption{Left: entanglement profiles for the single domain wall, for various $h$ and $\gamma=0.5$.
The parameters are $N=400$, $n_1=201$ and $t=200$. The solid red lines show the ansatz \eqref{entansatz}.
Right: half-chain entanglement as a function of time. The horizontal dotted line indicates the value $\ln 2$.
The inset shows the deviation from $\ln 2$ on a logarithmic scale.
The red dashed lines with slopes $-1/2$ and $-1$, respectively, are guides to the eye.}
\label{fig:entd}
\end{figure}

The entropy profiles for the single domain wall, located initially in the center
($r=0$) of the chain, have already been considered in \cite{EM18} and are shown
in the left of Fig. \ref{fig:entd} for $\gamma=0.5$ and several values of $h$.
The profile $\Delta S(r)=S(\rhoua)-S_0$ is always measured from
the initial entropy $S_0$ of the bulk ferromagnetic state, and is plotted against
the rescaled distance $\zeta=r/t$ from the center of the chain.
The main feature to be seen is the
emergence of a kink in the profile for $h<h_c$, at the value $\zeta_*$ that
equals the local maximum of the quasiparticle velocity, in complete analogy
to the case of the magnetization.

%
%

Due to the similar features observed in the entropy and magnetization profiles,
one is naturally led to the question whether there is a simple relation between
the two of them. We are also motivated by recent results of Refs. \cite{CADFDS18a,CADFDS18b},
where the entanglement content of particle excitations in $1+1$-dimensional massive quantum
field theories was studied, with a surprisingly simple result.
Namely, it has been found that the entropy difference (relative to the ground state)
of a single-mode excitation is independent of the wavenumber and given by
the binary entropy formula involving the ratio of the subsystem and full system lengths
\cite{CADFDS18a,CADFDS18b}. This ratio is just the density fraction of the single-mode
excitation that is contained within the subsystem.

Inspired by these findings, we put forward the following ansatz
\eq{
\Delta S(\zeta) = - \mathcal{N} \ln \, \mathcal{N} - 
(1-\mathcal{N}) \ln \, (1- \mathcal{N}) \, , \qquad
\mathcal{N}(\zeta) = \int_{-\pi}^{\pi} \frac{\dd P}{2\pi}  \Theta(v_P - \zeta) \, .
\label{entansatz}}
In other words, we assume that the static results of \cite{CADFDS18a,CADFDS18b}
would generalize to our dynamical scenario, and the entropy difference for bipartitions
along the ray $\zeta$ is just given by the same binary formula, with the density
ratio $\mathcal{N}(\zeta)$ being the fraction of the quasiparticles that have reached
the entangling point. Surprisingly, we find that the simple-minded ansatz \eqref{entansatz},
shown by the red solid lines in the left of Fig. \ref{fig:entd}, gives a very good
description of the entropy profiles. Via the density fraction $\mathcal{N}(\zeta)$,
the entropy profiles are thus directly related to those of the magnetization \eqref{mntsc}.

In case $h < h_c$, one observes some deviations from the ansatz
\eqref{entansatz}, which are only visible in the regime $\zeta <  \zeta_*$
and are assumed to be finite-time effects.
In order to better understand the convergence, on the right of Fig. \ref{fig:entd}
we also studied the time evolution of the half-chain entropy $\Delta S(0)$,
for the same parameter values.
Although each of them can be seen to converge towards the asymptotic
value $\ln 2$, their approach is rather different. For $h>h_c$ the convergence
is fast and steady, with rapid oscillations only, whereas for $h<h_c$ there is a
smaller frequency appearing with a larger amplitude, and the curve bounces back
from its asymptotical value repeatedly.
Interestingly, at the critical point $h=h_c=0.75$ one can see a slowing down
in the convergence, which becomes most evident on a logarithmic scale as
shown on the inset of the figure. Indeed, the approach seems to be
a power law $t^{-1/2}$, as opposed to $t^{-1}$ in the $h \ne h_c$ case.
This critical slowing down is responsible for the dip around $\zeta=0$ in the
profile for $h=h_c$ on the left of Fig. \ref{fig:entd}.

One should stress the marked difference of the entropy profiles as
compared to domain-wall evolution in critical systems, such as the XX chain.
Indeed, in the latter case the entropy was found to grow logarithmically in time
in the entire front region \cite{EP14,DSVC17}, whereas here the profiles 
converge to the scaling function \eqref{entansatz} when plotted against $\zeta=r/t$.
In particular, the result $\Delta S(0) = \ln 2$ for $\zeta=0$ implies that the entropy
converges towards the value attained in the ground state $\nsr{0}$, which has been
studied in \cite{Peschel04,IJK05}. Indeed, applying the relation \eqref{sup}
at $t=0$, one finds that the entropy $S_0$ in the initial symmetry-broken ground state
is exactly $\ln 2$ less than that of the NS ground state. This strongly suggests that the steady state is
nothing but the ground state with its symmetry restored.

\subsection{Double domain wall}

The profiles for the double domain wall are shown in Fig. \ref{fig:entdd} for
various times and two different model parameters.
In both cases, the profiles resemble those of two separate single domain walls
for short times, while for large times the main feature is the emergence of
an additional plateau in the overlap region. This strongly suggests the relation
\eq{
\Delta S_{n_1,n_2}(r) = \Delta S_{n_1}(r) + \Delta S_{n_2}(r) \, ,
\label{entadd}}
where $\Delta S_{n_1,n_2}(r)$ and $\Delta S_{n_i}(r)$ denote the
entropy differences for double and single domain walls, respectively,
with the indices referring to the initial locations of the excitations.
In other words, one expects the entropy differences to be additive,
which is indeed perfectly confirmed by the numerics.

%
\begin{figure}[htb]
\center
\includegraphics[width=0.49\columnwidth]{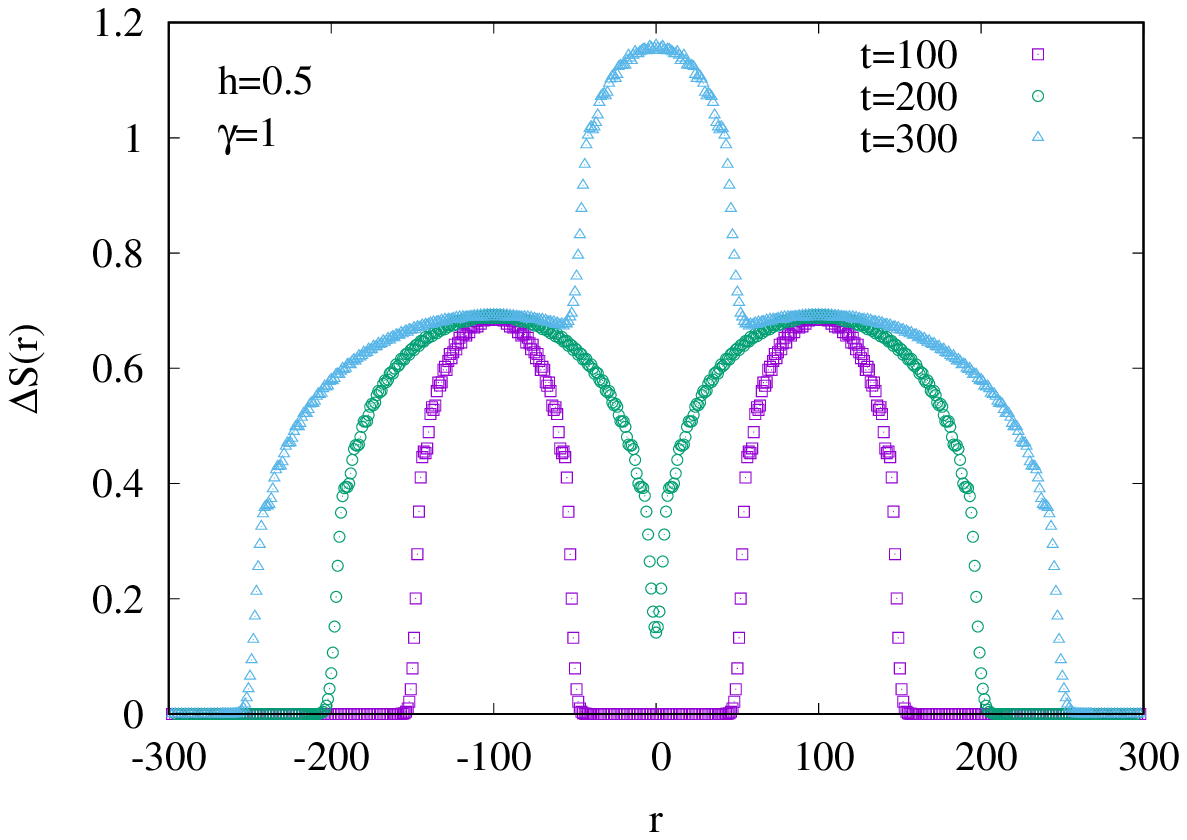}
\includegraphics[width=0.49\columnwidth]{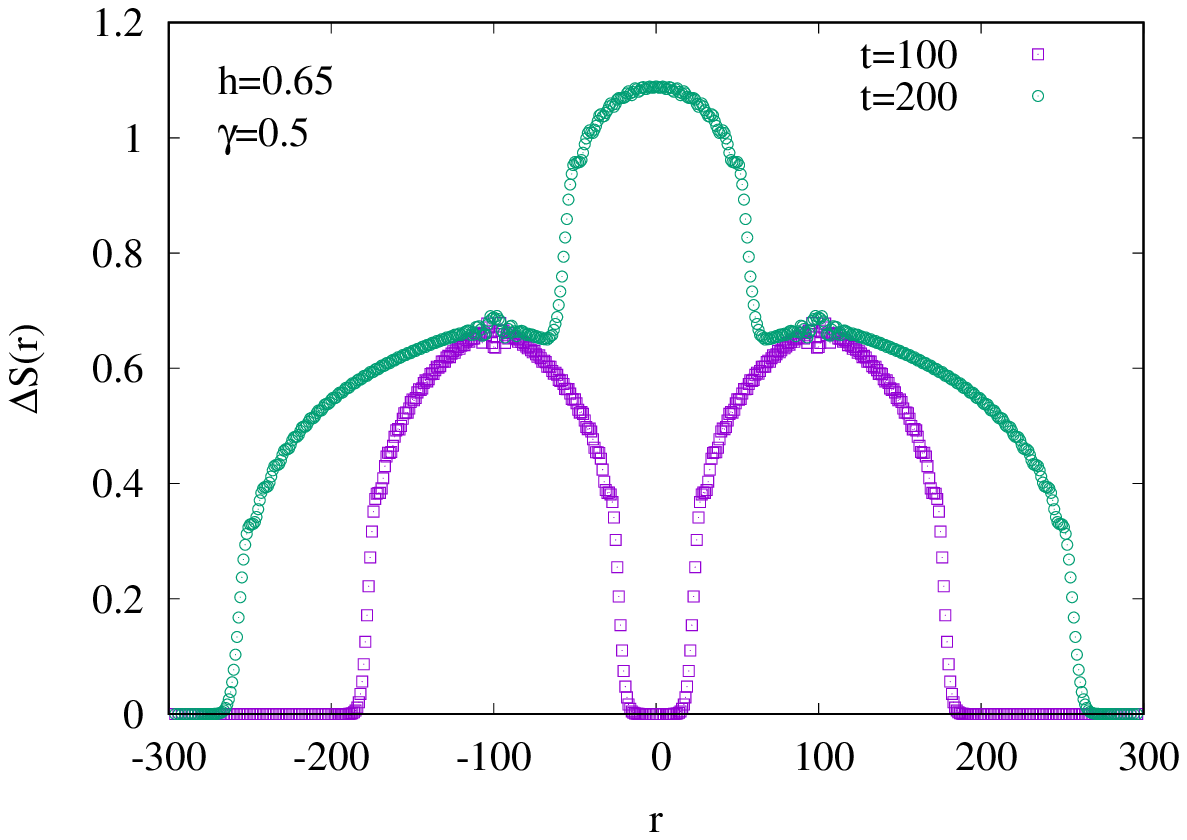}
\caption{Entanglement profiles after the double domain
wall excitation for different $h$ and $\gamma$. The parameters are the same
as in Fig. \ref{fig:mntdd}.}
\label{fig:entdd}
\end{figure}
%

\subsection{Single spin-flip}

Finally, we consider the entropy profiles for the spin-flip excitation, with the results
shown in Fig. \ref{fig:entdd}, for the same choice of parameters as for the
magnetization profiles in Fig. \ref{fig:mntsz}. When plotted against
the scaling variable $\zeta$, the profiles show a different behaviour as compared
to those of the single domain wall excitation in Fig. \ref{fig:entd}.
In particular, the additivity \eqref{entadd} is not satisfied, 
analogously to the corresponding result \eqref{mntsz} for the magnetization,
which does not have a factorized form. Indeed, as explained under Sec.
\ref{sec:magsz}, this has to do with an interference effect in the dynamics,
where an incoming momentum of the first excitation can travel forward as
an outgoing momentum of the second one. Clearly, such a process
creates entanglement between the quasiparticles building up the
two domain-wall excitations, which spoils the additivity and reduces the
overall entropy of the state. Unfortunately, despite the qualitative understanding
of the origin of the nontrivial entropy behaviour, we have not been able
to find an ansatz analogous to \eqref{entansatz} that captures the profiles
quantitatively.

%
\begin{figure}[htb]
\center
\includegraphics[width=0.49\columnwidth]{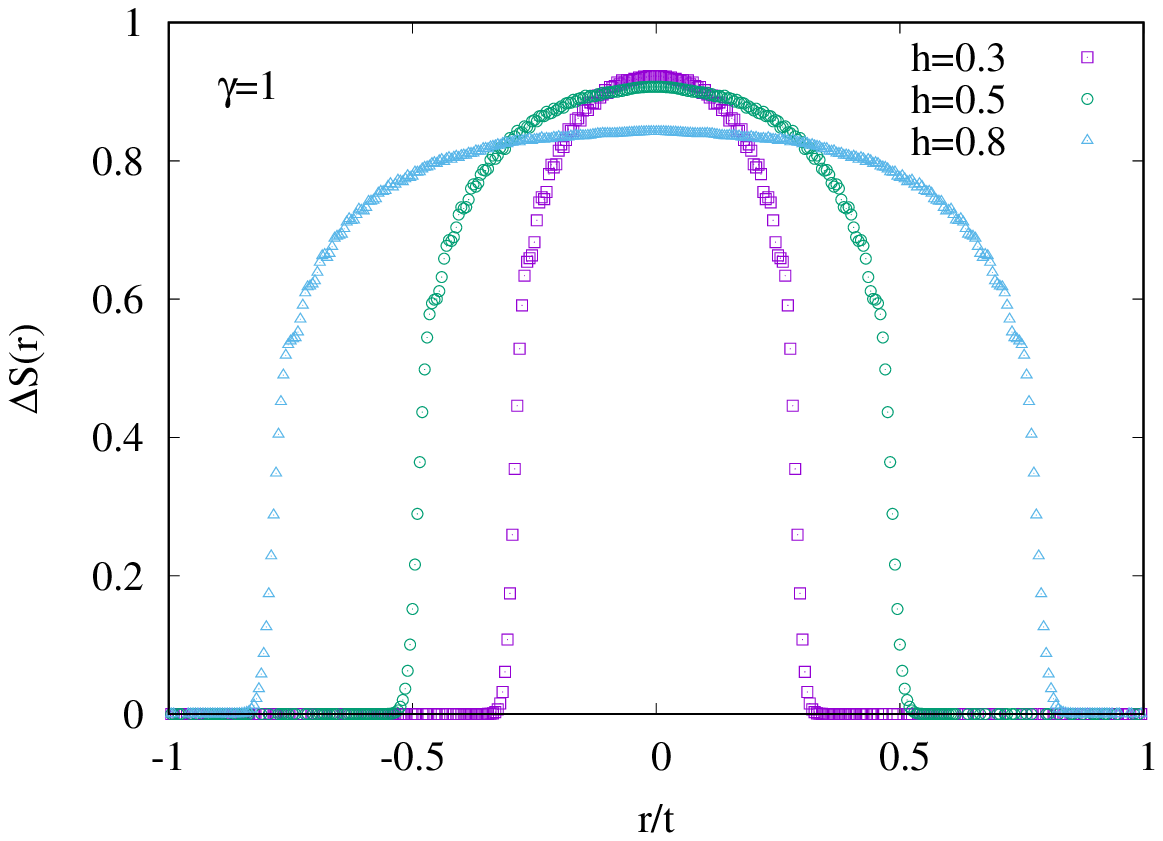}
\includegraphics[width=0.49\columnwidth]{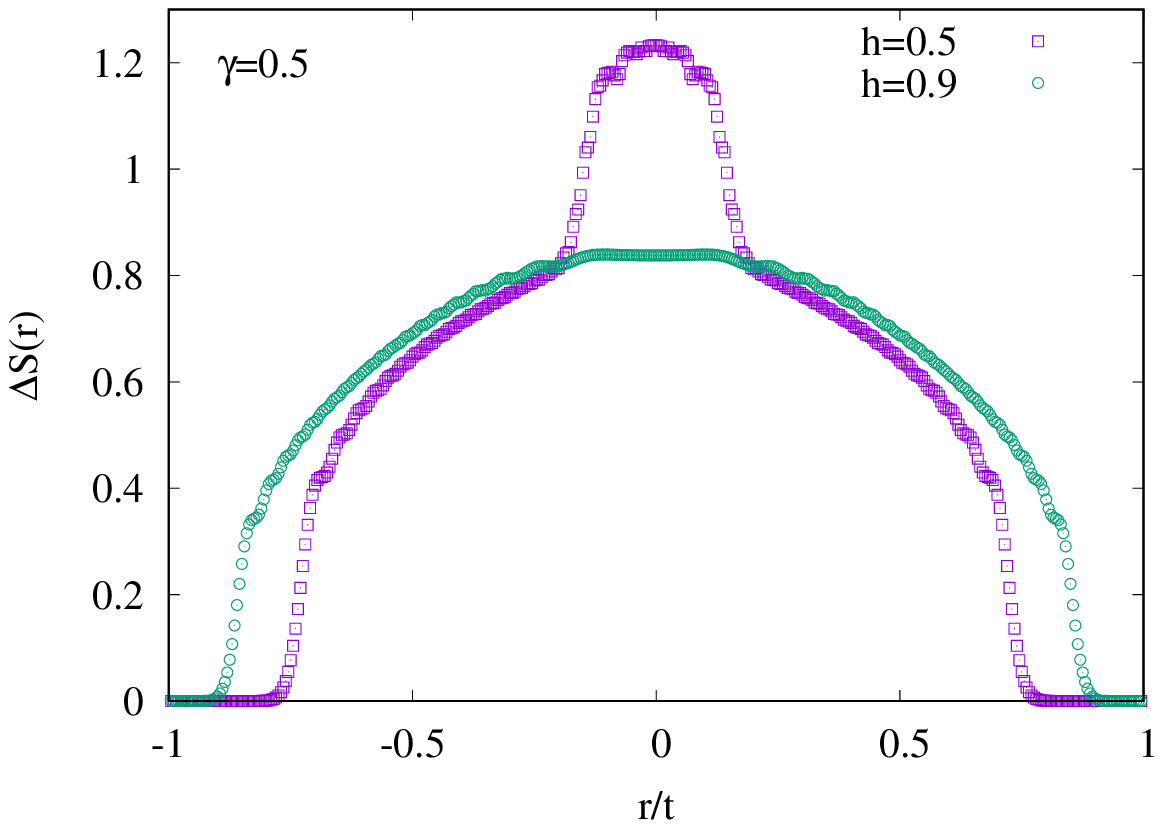}
\caption{Entanglement profiles after a spin-flip excitation.
The parameters are the same as in Fig. \ref{fig:mntsz}.
}
\label{fig:entdd}
\end{figure}
%

\section{Discussion\label{sec:disc}}

We studied the time evolution of the magnetization and entanglement
profiles in the XY chain for simple initial states that can be written as a
product of one or two local fermionic excitations. The former corresponds
to a single domain wall in the spin-picture and the magnetization
profile has a simple hydrodynamic limit \eqref{mntsc}, corresponding
to the motion of independent quasiparticles.
Furthermore, in the very same limit we find that the entropy is given by
the simple ansatz \eqref{entansatz} and is thus directly related to the
magnetization profile. The correlation function is also found to be
related via \eqref{cmn} to the magnetization, which gives a very
good approximation even for finite times and distances.

For double domain walls, excited by the product
of two fermions separated by a large distance, we simply find the
factorized form \eqref{mntdd} for the magnetization, as well as the
additivity \eqref{entadd} of the entropy differences.
For a single spin-flip, however, the fermions are located on neighbouring
sites and the excitation cannot be considered strictly local any more.
As a consequence, we find an interference term in the magnetization
profile \eqref{mntsz}.
Furthermore, the additivity of the entropy is lost, and we find convergence
towards a nontrivial profile

We have also compared the profiles for the single domain wall to the ones
obtained via a local cut and glue quench, where the two ferromagnetic
ground states are prepared on half-chains and joined together.
Rather surprisingly we found that, while for the TI chain the respective
profiles almost coincide, for the generic XY case they become completely
different (see Fig. \ref{fig:mlq}), with the discrepancy growing with the distance
from the TI limit. Apparently the local quench is well approximated by a
single fermionic excitation for the TI but not any more for the XY case.
The precise origin of this phenomenon is unclear to us and requires
further studies.

It would be also interesting to see if a QFT treatment of the entropy increase
could be given. Even though our ansatz \eqref{entansatz} was inspired by
QFT results \cite{CADFDS18a,CADFDS18b} on the entanglement
content of particle excitations, those particles are single wave modes
and there is no dynamics involved. On the other hand, for the case of critical
Hamiltonians there exists a CFT framework for calculating the time evolution of
entropy after spatially local excitations \cite{NNT14}. Whether this approach could
be generalized to a massive QFT to predict the asymptotic entropy increase after
the excitations considered in this paper is a puzzling question to be addressed.

One could also think about extending the studies to excitations composed of
a product of more than two fermions. While being a straightforward generalization,
the form-factor calculations are likely to be very cumbersome, due to the
increasing number of the pole contributions one has to account for.
Finally, it is natural to ask how the setup could be extended to interacting
integrable systems, and if the treatment of such composite but still essentially
local excitations could be incorporated into the theory of GHD.

\section*{Acknowledgements}

We thank H. G. Evertz for discussions and collaboration on a related
previous project.

\paragraph{Funding information}
The authors acknowledge funding from the Austrian Science Fund (FWF) through
Project No. P30616-N36, and through SFB ViCoM F41 (Project P04).

\begin{appendix}

\section{Form factors for the TI and XY chains\label{app:ff}}

Here we present the form factors used in the calculations of the main text.
Although for our simple excitations we required only few-particle form factors,
the general expression is reported for completeness.
The formula is rather involved even after taking the thermodynamic limit
$N\to\infty$, and for the TI chain ($\gamma=1$) it reads \cite{IST11}
\begin{align}
\frac{\rl{p_1,\dots,p_L} \sigma^x_n \nsr{q_1,\dots,q_K}} {\rl{0} \sigma^x_n \nsr{0}} =
i^{-(K+L)/2} \, (-1)^{L(L-1)/2} \, h^{(K-L)^2/4} \, 
\ee^{in(\sum_{k=1}^{K} q_k - \sum_{l=1}^{L} p_l)} \nonumber \\  \times
\prod_{k=1}^{K} \frac{1}{\sqrt{N \epsilon_{q_k}}}
\prod_{l=1}^{L} \frac{1}{\sqrt{N \epsilon_{p_l}}}
\prod_{k<k'=1}^{K}\frac{\sin \frac{q_k-q_{k'}}{2}}{\frac{\epsilon_{q_k}+\epsilon_{q_{k'}}}{2}}
\prod_{l<l'=1}^{L}\frac{\sin \frac{p_l-p_{l'}}{2}}{\frac{\epsilon_{p_l}+\epsilon_{p_{l'}}}{2}}
\prod_{k=1}^{K}\prod_{l=1}^{L}
\frac{\frac{\epsilon_{q_k}+\epsilon_{p_l}}{2}}{\sin \frac{q_k-p_l}{2}} \, .
\label{ffti}
\end{align}
We have assumed here that the number of momenta $K$ and $L$ on the
right and left hand side have the same parity, otherwise the form factor vanishes.
Note that we have normalized with the vacuum form factor, i.e. with the expectation 
value of the ground-state magnetization. For $K=L$ the form factors \eqref{ffti} depend
only on the dispersion relation $\epsilon_q$, given in Eq. \eqref{epsq}, and the values
of the momenta.

For the generic case of the XY chain, the expressions become even more complicated.
In the limit $N\to\infty$ they can be written as \cite{Iorgov11,IL11}
%
\begin{align}
&\frac{\rl{p_1,\dots,p_L} \sigma^x_n \nsr{q_1,\dots,q_K}} {\rl{0} \sigma^x_n \nsr{0}} =
i^{-(K+L)/2} \, (-1)^{L(L-1)/2} \, g^{(K-L)^2/4} \, 
\ee^{in(\sum_{k=1}^{K} q_k - \sum_{l=1}^{L} p_l)}
\nonumber \\  & \qquad\qquad \times
\cosh \frac{\sum_{k=1}^{K} \Delta_{q_k} - \sum_{l=1}^{L} \Delta_{p_l}}{2}
\prod_{k=1}^{K} \frac{1}{\sqrt{N \sinh\Delta_{q_k}}}
\prod_{l=1}^{L} \frac{1}{\sqrt{N \sinh\Delta_{p_l}}}
\nonumber \\  & \qquad\qquad \times
\prod_{k<k'=1}^{K}\frac{\sin \frac{q_k-q_{k'}}{2}}{\sinh \frac{\Delta_{q_k}+\Delta_{q_{k'}}}{2}}
\prod_{l<l'=1}^{L}\frac{\sin \frac{p_l-p_{l'}}{2}}{\sinh \frac{\Delta_{p_l}+\Delta_{p_{l'}}}{2}}
\prod_{k=1}^{K}\prod_{l=1}^{L}
\frac{\sinh \frac{\Delta_{q_k}+\Delta_{p_l}}{2}}{\sin \frac{q_k-p_l}{2}} \, ,
\label{ffxy}
\end{align}
where we have defined
\eq{
\sinh \Delta_q = 
\frac{\sqrt{1-\gamma^2}}{\gamma \sqrt{\gamma^2+h^2-1}} \epsilon_q \, , \qquad
g = \frac{1-\gamma^2}{\gamma \sqrt{\gamma^2+h^2-1}} \, .
\label{shd}}
The above definition is valid in the parameter regime $\sqrt{1-\gamma^2}<h<1$,
i.e. in the non-oscillatory ferromagnetic phase.
In the oscillatory phase $0<h< \sqrt{1-\gamma^2}$ the corresponding expressions
can be obtained by analytic continuation \cite{Iorgov11}.
One can also check that, in the singular TI limit $\gamma \to 1$, the expression
\eqref{ffxy} goes over to the one in \eqref{ffti}. While in general they differ in the
details, these will turn out to be irrelevant for the hydrodynamic limit, since their
pole structure is exactly the same.

We now discuss the form factors needed in the main text. The simplest
is the one-particle form factor ($K=L=1$), where using some hyperbolic
identities in \eqref{ffxy}, one can show that the TI and XY cases yield
the same expression
\eq{
\frac{\rl{p} \sigma^x_n \nsr{q}}{\rl{0} \sigma^x_n \nsr{0}} =
-\frac{i}{N}\frac{\epsilon_p + \epsilon_q}{2\sqrt{\epsilon_p \epsilon_q}}
\frac{\ee^{i n(q-p)}}{\sin \frac{q-p}{2}}.
\label{ff11}}
Thus the formula \eqref{mntd} for the single domain wall excitation is valid 
for arbitrary parameter values of the XY chain. In general, no such simplification
occurs and in the following we restrict ourselves to the TI case for the sake
of simplicity. For the spin-flip excitation one needs the off-diagonal form factor
with $K=2$ and $L=0$ which reads
\eq{
\frac{\rl{0} \sigma^x_n \nsr{q_1,q_2}} {\rl{0} \sigma^x_n \nsr{0}} =
-\frac{i}{N} \frac{h}{\sqrt{\epsilon_{q_1}\epsilon_{q_2}}} \ee^{in(q_1+q_2)}
\frac{2\sin \frac{q_1-q_2}{2}}{\epsilon_{q_1}+\epsilon_{q_2}}.
\label{ff20}}
One can see immediately, that this form factor does not have any poles
which implies that it will only give a subleading contribution.
The diagonal two-particle form factor ($K=L=2$), on the other hand,
has the form
\begin{align}
\frac{\rl{p_1,p_2} \sigma^x_n \nsr{q_1,q_2}} {\rl{0} \sigma^x_n \nsr{0}} &=
 \frac{1}{N^2}
\frac{\ee^{in(q_1+q_2-p_1-p_2)}}{\sqrt{\epsilon_{p_1}\epsilon_{p_2}\epsilon_{q_1}\epsilon_{q_2}}} 
\frac{2\sin \frac{p_1-p_2}{2}}{\epsilon_{p_1}+\epsilon_{p_2}}
\frac{2\sin \frac{q_1-q_2}{2}}{\epsilon_{q_1}+\epsilon_{q_2}}
\nonumber \\  &\times 
\frac{\epsilon_{q_1}+\epsilon_{p_1}}{2\sin \frac{q_1-p_1}{2}}
\frac{\epsilon_{q_1}+\epsilon_{p_2}}{2\sin \frac{q_1-p_2}{2}}
\frac{\epsilon_{q_2}+\epsilon_{p_1}}{2\sin \frac{q_2-p_1}{2}}
\frac{\epsilon_{q_2}+\epsilon_{p_2}}{2\sin \frac{q_2-p_2}{2}} \, ,
\label{ff22}
\end{align}
with two possible poles for $q_1=p_1$ and $q_2=p_2$,
or with an exchange of momenta for $q_1=p_2$ and $q_2=p_1$.
It should be noted that, for the generic diagonal $K$-particle form factors in \eqref{ffti},
an arbitrary permutation between the incoming and outgoing momenta yields a pole,
which makes the analysis of the contributions increasingly complicated.

\section{Stationary phase calculations for the profile\label{app:sp}}

In this appendix we summarize the calculations leading to the approximations
of the magnetization profile in the hydrodynamic regime. The simplest case is
the single domain wall, where $\mnt$ is given by a double integral \eqref{mntd}.
The integrand has a pole due to the form factor, which can be regularized as
\eq{
\mnt= 1 +
\int_{-\pi} ^{\pi} \frac{\dd p}{2\pi} \int_{-\pi}^{\pi} \frac{\dd q}{2\pi}
\frac{\epsilon_p + \epsilon_q}{2\sqrt{\epsilon_p \epsilon_q}}
\frac{\ee^{i(n-n_1+1/2)(q-p)}}{i\sin\left(\frac{q-p+i\varepsilon}{2}\right)}
\ee^{i(\theta_q - \theta_p)/2} \ee^{-i(\epsilon_q-\epsilon_p)t} \, ,
\label{mnteps}}
by introducing the infinitesimal shift $\varepsilon>0$.
The integrand of \eqref{mnteps} is highly oscillatory for $|n-n_1|\gg 1$ and $t \gg 1$,
and the location of the pole at $q=p$ suggests the change of variables
$Q=q-p$ and $P=(q+p)/2$. The phase factors become stationary at $Q=0$,
thus the integrand should be expanded around this value. Keeping only the
most singular term one has
\eq{
1 + 2 \, \int_{-\pi}^{\pi} \frac{\dd P}{2\pi}\int_{-\infty}^{\infty} \frac{\dd Q}{2\pi i}
\frac{\ee^{i(n-n_1+1/2+\theta'_{P}-v_{P}t)Q}}{Q+i\varepsilon} \, ,
}
where we have extended the integration in the relative momentum up to infinity.
Thanks to the definition \eqref{epsq}, the function $\theta'_{P}$ varies smoothly and
one can neglect it in the hydrodynamic regime. Then using the integral representation
of the Heaviside theta function
\eq{
\Theta(x) = -\lim_{\varepsilon \to 0} \int_{-\infty}^{\infty} \frac{\dd Q}{2\pi i}
\frac{\ee^{-iQx}}{Q+i\varepsilon},
\label{ht}
}
and introducing the ray variable $\nu=(n-n_1+1/2)/t$ brings us 
to the result \eqref{mntsc} in the main text.

The bulk hydrodynamic profile is thus recovered by solving the equation $v_q=\nu$.
Special attention is needed around the maximum $v_{q_*}=v_{max}$ of the velocities,
where the solutions coalesce at momentum $q_*$. To get the fine structure of the
front edge, one has to expand the dispersion around $q_*$ as
\eq{
\epsilon_q \approx \epsilon_{q_*} + v_{q_*} (q-q_*) +
\frac{\epsilon'''_{q_*}}{6} (q- q_*)^3.
\label{epsqmax}}
Furthermore, one can introduce the following rescaled variables
\eq{
\begin{split}
X = \left(\frac{-2}{\epsilon'''_{q_*}t}\right)^{1/3} (n-n_1+1/2+\theta'_{q_*}/2-v_{q_*}t), \\
Q = \left(\frac{-2}{\epsilon'''_{q_*}t}\right)^{-1/3} (q-q_*), \quad
P = \left(\frac{-2}{\epsilon'''_{q_*}t}\right)^{-1/3} (p-q_*).
\end{split}
\label{XQP}}
Substituting \eqref{epsqmax} and \eqref{XQP} into \eqref{mnteps},
one arrives at the following integral
\eq{
1 + 2 \left(\frac{-2}{\epsilon'''_{q_*}t}\right)^{1/3}
\int \frac{\dd P}{2\pi} \int \frac{\dd Q}{2\pi}
\frac{\ee^{iX(Q-P)} \ee^{i(Q^3-P^3)/3}}{i(Q-P+i\varepsilon)}.
}
Using the integral representation of the Airy kernel \cite{TW94}
\eq{
\mathcal{K}_{Ai}(X,Y) = \lim_{\varepsilon\to0} \int \frac{\dd P}{2\pi} \int \frac{\dd Q}{2\pi}
\frac{\ee^{-iXP} \ee^{-iP^3/3} \ee^{iYQ} \ee^{iQ^3/3}}{i(P-Q-i\varepsilon)}=
\frac{\Ai(X)\Ai'(Y)-\Ai'(X)\Ai(Y)}{X-Y},
\label{kxy}}
one recovers \eqref{medge} of the main text, with
$\rho(X)=\lim_{Y \to X} \mathcal{K}_{Ai}(X,Y)$
given by the diagonal terms of the Airy kernel.

The hydrodynamic limit \eqref{mntdd} for the double domain wall can
be obtained in a similar fashion, however, one has now a quadruple
integral to start with. The poles are contained in the two-particle form
factor \eqref{ff22}. First, we consider the pole with $q_1=p_1$ and
$q_2=p_2$. Changing variables as
\eq{
Q_i = q_i-p_i, \qquad P_i = \frac{q_i+p_i}{2},
\label{QK}}
and expanding the phases around the stationary points $Q_i=0$,
one has
\eq{
\mathcal{I}_1=4\int \frac{\dd P_1}{2\pi} \int \frac{\dd P_2}{2\pi} 
f(P_1,P_2,Q_1,Q_2)
\int \frac{\dd Q_1}{2\pi} \frac{\ee^{-ix_1Q_1}}{Q_1}
\int \frac{\dd Q_2}{2\pi} \frac{\ee^{-ix_2Q_2}}{Q_2} \, ,
\label{intp1}}
where we defined
\eq{
x_i = v_{P_i}t - (-1)^i \theta'_{P_i} - (n-n_i+1/2) \, .
}
The function $f$ in \eqref{intp1} describes the slowly varying part of the
form factor in \eqref{ff22}. It is easy to see, that the terms containing the
dispersion $\epsilon_{q_i}$ and $\epsilon_{p_i}$ can be approximated
by $1$ to leading order. It remains to analyze the contribution of the
trigonometric factors that do not contain the poles, which can be rewritten as
\eq{
f(P_1,P_2,Q_1,Q_2) \approx
- \frac{\cos(\frac{Q_1-Q_2}{2})-\cos(P_1-P_2)}{\cos(\frac{Q_1+Q_2}{2})-\cos(P_1-P_2)}.
}
Thus, again to leading order around $Q_i=0$, one has
$f(P_1,P_2,Q_1,Q_2) \approx -1+\mathcal{O}(Q_1Q_2)$, meaning
that the first correction would already remove the singularity in the integral
\eqref{intp1}, and can be neglected. Setting $f = -1$, one recovers
immediately the factorized result \eqref{mntdd}.

The second pole of the form factor \eqref{ff22} is given by
$q_1=p_2$ and $q_2=p_1$ and corresponds to an exchange of the
outgoing momenta. The form factor itself transforms trivially under this
exchange, acquiring only a sign. The time-evolved state \eqref{ddt},
however, has phase factors attached to the locations of the domain walls
and thus transforms nontrivially under exchange of the momenta.
Indeed, introducing the variables
\eq{
Q'_1 = q_1-p_2, \qquad Q'_2 = q_2-p_1, \qquad
P'_1 = \frac{q_1+p_2}{2}, \qquad P'_2 = \frac{q_2+p_1}{2},
}
this phase factor can now be rewritten as
\eq{
\ee^{-i(Q'_1+Q'_2)(n_1+n_2)/2} \, 
\ee^{i(P'_1-P'_2)(n_2-n_1)} .
\label{phase}}
The second term contains the center of mass momenta and becomes
highly oscillatory for $|n_2-n_1| \gg 1$. This phase, however, cannot
be made stationary, since the time-dependent part of the phase
in \eqref{ddt} is symmetric under the exchange of the momenta.
One thus concludes that, for large separations of the domain walls,
the second pole gives a negligible contribution.

The situation for the spin-flip excitation is different. As discussed
in the main text, except for a sign change of the Bogoliubov angles,
the state \eqref{sz0} is a double domain wall with $n_2=n_1+1$.
The first pole thus yields the very same factorized result as in \eqref{intp1},
with the corresponding changes in $x_i$. In the hydrodynamic limit,
however, it is more natural to measure distances from the spin-flip location $n_1$
(instead of $n_1\pm 1/2$) and use the ray variable $\tilde \nu=\frac{n-n_1}{t}$,
which gives the second term in \eqref{mntsz}.
The second pole, however, has also a significant contribution, since $n_2-n_1=1$
and the phase factor in \eqref{phase} now varies slowly.
Expanding around $Q'_i=0$, one finds
\eq{
\mathcal{I}_2=4\int \frac{\dd P'_1}{2\pi} \int \frac{\dd P'_2}{2\pi}
\ee^{iP'_1}\ee^{i\theta_{P'_1}}\ee^{-iP'_2}\ee^{-i\theta_{P'_2}}
\int \frac{\dd Q'_1}{2\pi} \frac{\ee^{-ix'_1Q'_1}}{Q'_1}
\int \frac{\dd Q'_2}{2\pi} \frac{\ee^{-ix'_2Q'_2}}{Q'_2}\, ,
}
where $x'_i = v_{P'_i}t - (n-n_1)$ and the sign change in the form factor
has been taken into account. It is easy to see that
\eq{
\mathcal{I}_2 = -\left|
2\int \frac{\dd P'}{2\pi} \ee^{iP'}\ee^{i\theta_{P'}}
\int \frac{\dd Q'}{2\pi i} \frac{\ee^{-ix'Q'}}{Q'}\right|^2 .
}
Regularizing the pole via the identity
$Q'^{-1}=i \pi \delta(Q') + \lim_{\varepsilon \to 0}(Q'+i\varepsilon)^{-1}$,
using \eqref{ht} and the expression of the transverse magnetization in \eqref{mz},
the third term of \eqref{mntsz} follows.

It remains to investigate the edge scaling regime for the spin-flip excitation.
The second term of \eqref{mntsz} is simply the square of the profile for a single domain
wall, where the edge scaling is given by \eqref{medge}. To leading order, this just
yields a factor 2. The situation is similar for the third term in \eqref{mntsz}
where, additionally, the phase factors in the integral must be evaluated at the
momentum $q_*$ where the velocity has its maximum,  $v_{q_*}=v_{max}$.
Collecting the terms, one obtains the prefactor in \eqref{rhot}.

Finally it should be noted that, although the calculation above has been carried
out using the form factors for the TI chain, the result generalizes to the XY case.
Indeed, the pole structure of the form factors is exactly the same, whereas the
differences in the slowly varying part are irrelevant in the hydrodynamic limit,
since they have the same trivial limit after expanding around the pole.

\section{Calculation of correlation functions\label{app:cf}}

At one-particle level of the form-factor expansion, the normalized correlation
function is given by the triple integral
\begin{align}
\cmnt \simeq
&\int \frac{\dd q_1}{2\pi}\int \frac{\dd q_2}{2\pi}
\ee^{-i(\theta_{q_1}-\theta_{q_2})/2}\ee^{i(\epsilon_{q_1} -\epsilon_{q_2})t}
\nonumber \\  \times
&\int \frac{\dd p}{2\pi} \frac{\epsilon_p + \epsilon_{q_1}}{2\sqrt{\epsilon_p \epsilon_{q_1}}}
\frac{\epsilon_p + \epsilon_{q_2}}{2\sqrt{\epsilon_p \epsilon_{q_2}}}
\frac{\ee^{-i(m-n_1+1/2)(q_1-p)}}{\sin \frac{q_1-p}{2}}\frac{\ee^{i(n-n_1+1/2)(q_2-p)}}{\sin \frac{q_2-p}{2}} \, .
\label{cmnintapp}
\end{align}
The stationary phase approximation of this integral is very similar to that
of the magnetization profile. Introducing the new set of variables
\eq{
Q_1 = q_1-p, \qquad
Q_2 = q_2-p, \qquad
P = \frac{q_1+p}{2},
}
and expanding around the poles $Q_1=0$ and $Q_2=0$, one obtains
\eq{
\cmnt \simeq
4\int \frac{\dd P}{2\pi}
\int \frac{\dd Q_1}{2\pi}
\frac{\ee^{-i(m-n_1+1/2 + \theta'_P -v_Pt)Q_1}}{Q_1}
\int \frac{\dd Q_2}{2\pi}
\frac{\ee^{i(n-n_1+1/2 + \theta'_P -v_Pt)Q_2}}{Q_2}.
\label{cxint2}}
Applying \eqref{ht} in both the $Q_1$ and $Q_2$ integrals, the result can
again be written with the help of step functions
\eq{
\cmnt \simeq 
1 - 2\int_{-\pi}^{\pi} \frac{\dd P}{2\pi}
\left[ \Theta(v_{P} - \mu) + \Theta(v_{P} - \nu) -
2 \Theta(v_{P} - \mu)\Theta(v_{P} - \nu) \right],
}
where the scaling variable $\mu =(m-n_1+1/2)/t$ is introduced analogously to $\nu$.
Assuming $\mu<\nu$ and using the identities for the step function
\eq{
\Theta(v_{P} - \nu) = 1 - \Theta(\nu - v_{P}) \, , \qquad
\Theta(v_{P} - \mu) - \Theta(v_{P} - \nu) = \Theta(v_{P} - \mu) \Theta(\nu - v_{P}) \, ,
}
the result \eqref{cmnsc} of the main text follows immediately.

Instead of applying the stationary phase argument, one can also do a more precise
analysis. Indeed, it turns out that the integral over $p$ in \eqref{cmnintapp} can be
carried out explicitly. We first regularize the factor containing the pole as
\eq{
\frac{1}{\sin \left( \frac{q_1-p}{2} \right) \sin \left( \frac{q_2-p}{2} \right)}=
\left[2\pi \delta(p-q_1) + \frac{1}{i\sin\left( \frac{q_1-p+i\varepsilon}{2} \right)}\right]
\left[2\pi \delta(p-q_2) + \frac{1}{i\sin\left( \frac{p- q_2+i\varepsilon}{2} \right)}\right] .
\label{poles}}
Multiplying out this expression, the terms containing the delta functions can
be plugged back into \eqref{cmnintapp} and integrated over.
Comparing to \eqref{mnteps}, one can identify the resulting double integrals as
$\mathcal{M}_{m}(t)-1$ and $\mathcal{M}_{n}(t) - 1$, respectively, while
the product of the delta functions trivially yields one.
The remaining factor from \eqref{poles} can be rewritten as
\eq{
\frac{1}{\sin \left( \frac{q_1-p+i\varepsilon}{2} \right) \sin \left( \frac{p-q_2+i\varepsilon}{2} \right)}=
\frac{2}{\cos \left( \frac{q_1+q_2}{2}-p \right) -
\cos \left( \frac{q_1-q_2}{2}+i\varepsilon \right)} \, .
}
Introducing new variables
\eq{
z = \ee^{i\left[p-(q_1+q_2)/2\right]}\, , \qquad z_0 = \ee^{i\left[(q_1-q_2)/2+i\varepsilon\right]}\, ,
}
the integral over $p$ is transformed into the contour integral
\eq{
\mathcal{I}= \oint \frac{\dd z}{2\pi i z} f(z)
\frac{4}{z+ z^{-1} - (z_0+z_0^{-1})},
}
where $f(z)$ is the slowly varying regular part of the integrand in \eqref{cmnintapp},
and the contour is the unit circle.
Now the two poles are located at $z = z_0$ and $z=z_0^{-1}$.
However, for $\varepsilon>0$, only $z = z_0$ lies inside the contour and contributes
to the integral. We have thus to obtain the residue around this pole.
Rewriting
\eq{
\frac{4}{z^2 + 1-z(z_0+z_0^{-1})} =
\frac{4}{z_0-z_0^{-1}}
\left( \frac{1}{z-z_0} - \frac{1}{z-z_0^{-1}} \right),
}
and the two poles correspond to $p=q_1$ and $p=q_2$, respectively.
Hence the result of the contour integral is
\eq{
\mathcal{I} = \frac{2f(q_1)}{i \sin\left(\frac{q_1-q_2}{2}+i\varepsilon\right)} \, .
}
Finally, noting that $\mathcal{I}$ enters with a minus sign (see \eqref{poles}),
and inserting the result back into \eqref{cmnintapp}, one can easily identify the
contribution as $-2(\mathcal{M}_{n}(t) - 1)$. Collecting all the terms,
one arrives at the result
\eq{
\cmnt \simeq 1 + \mathcal{M}_{m}(t)-\mathcal{M}_{n}(t) \, .
\label{cvsm1p}}

As a closing remark, we give a simple argument why the many-particle contributions
in the form-factor expansion of the correlation functions can be neglected.
In the one-particle expression \eqref{cmnintapp}, the dominant contribution
is obtained from momenta satisfying $q_1=p=q_2$, where the stationary
phase conditions match the poles of the integrand. The next nonvanishing term
in the expansion involves three intermediate particles, where the phase factor could
be made stationary for $q_1=p_1=q_2$ and $p_2=-p_3$. However, from \eqref{ffti}
one can see that there is no pole in the form factor at $p_2=-p_3$, and thus the
contribution is subleading.

\end{appendix}

\bibliography{xylocalex_refs}

\end{document}